\documentclass[twocolumn,trackchanges]{aastex63}
\usepackage{amsmath}
\usepackage{graphicx}
\usepackage{appendix}
\usepackage{tabularx}
\usepackage{booktabs}
\usepackage{comment}
\usepackage{xcolor}
\usepackage{hyperref}
\usepackage{lineno}

\newcommand{\h}{\mathbf{h}}
\newcommand{\Msun}{\ensuremath{M_\odot}}        
\newcommand{\Lsun}{\ensuremath{L_\odot}}        
\newcommand{\Rsun}{\ensuremath{R_\odot}}        
\newcommand{\Mzams}{\ensuremath{M_{\rm ZAMS}}}  
\newcommand{\Teff}{\ensuremath{T_{\rm eff}}}    

\newcommand{\code}[1]{\texttt{#1}}
\newcommand{\mesa}{\code{MESA}}
\newcommand{\MESA}{\mesa}

\newlength{\apjcolwidth}
\begin{document}

\title{Self-Consistent Nonlinear Classical Cepheid Pulsations During Stellar Evolution with \MESA}

\shorttitle{Cepheid pulsations with MESA}
\shortauthors{Farag et al.}

\author[0000-0002-5794-4286]{Ebraheem Farag}
\affiliation{Department of Astronomy, Yale University, New Haven, CT 06511, USA}
\affiliation{School of Earth and Space Exploration, Arizona State University, Tempe, AZ 85287, USA}

\author[0000-0003-4456-4863]{Earl P. Bellinger}
\affiliation{Department of Astronomy, Yale University, New Haven, CT 06511, USA}

\author[0000-0001-6631-2566]{Philip Mocz}
\affiliation{Center for Computational Astrophysics, Flatiron Institute, 162 5th Ave, New York, NY 10010, USA}

\author[0009-0001-6503-9841]{Selim Kalici}
\affiliation{Department of Astronomy, Yale University, New Haven, CT 06511, USA}

\author[0000-0001-7217-4884]{R. Smolec}
\affiliation{Nicolaus Copernicus Astronomical Centre, Polish Academy of Sciences, Bartycka 18, 00-716 Warszawa, Poland}

\author[0000-0003-0791-7594]{Shashi Kanbur}
\affiliation{Department of Physics, State University of New York at Oswego, Oswego, NY 13126, USA}

\author[0009-0007-8010-8562]{Kyra Bettwy}
\affiliation{Department of Astronomy, Yale University, New Haven, CT 06511, USA}

\author[0000-0001-8722-1436]{Christopher Lindsay}
\affiliation{Department of Astronomy, Yale University, New Haven, CT 06511, USA}


\correspondingauthor{Ebraheem Farag}
\email{ebraheem.farag@yale.edu}

\begin{abstract}
We extend the time-dependent convection treatment in \code{MESA} by introducing eddy-viscous damping. This software change brings \code{MESA-TDC} into closer alignment with the radial stellar pulsation framework of \code{MESA-RSP}. We demonstrate that the inclusion of the eddy viscosity in hydrodynamic stellar models remains stable on evolutionary timescales. We then present the first self-consistent integration of large-amplitude, nonlinear Classical Cepheid pulsations directly within a \code{MESA-star} evolutionary run, demonstrating that the time-dependent convection formalism implemented in \code{MESA-star} and the \code{MESA} radial stellar pulsation (RSP) module are physically identical. Starting from a 6~\Msun\ blue-loop stellar evolution model, we demonstrate evolving the entire stellar model through pulsations as well as pausing the evolution, excising the core, and remeshing the envelope to match the grid used by \code{MESA-RSP}. We compare the pulsation properties (e.g., period, light and radius curves, and growth rate) with a matched \code{MESA-RSP} run, and find reasonable agreement between the two modules. This unified approach eliminates the reliance on separate post-processing workflows and enables fully coupled evolution-pulsation simulations. This approach enables future studies of stellar pulsations with the inclusion of composition gradients, mass loss, or rotation. It also enables future studies of the $\epsilon$ mechanism as well as providing a physical source of viscosity for other science cases explored using \code{MESA}'s hydrodynamics solver. We have integrated these modifications into the \code{MESA-star} module, enabling open-source use by the community. 
\end{abstract}


\section{Introduction} \label{s.intro}

Among the very first variable stars discovered were $\eta$~Aquilae and $\delta$~Cephei \citep{pigott_1785_aa,Goodricke_1786_aa} and were dubbed Cepheid variables, a class of stars characterized by their continuous brightening and dimming. 
\citet{leavitt_1912_aa} found the luminosity of such stars to depend on the period of their variation, thereby enabling measurement of distances to galaxies in the local universe and leading to the discovery of the expanding universe \citep{hubble_1926_aa}. Classical Cepheids remain at the heart of extragalactic distance measurements \citep{baade_1956_aa,feast_1987_aa} and serve as the bridge between Galactic parallax distances and extragalactic scales. With the discovery of the accelerating expansion rate of the universe, characterizing the properties of the pulsations of these stars has become one of the primary systematics on which almost all extragalactic distance indicators depend \citep{madore_1991_aa,riess_1998_aa,freedman_2001_aa,turner_2002_aa,freedman_2010_aa,freedman_2012_aa,riess_2016_aa,riess_2019_aa,freedman_2021_aa,riess_2022_aa}. 

More than a century after their discovery, \citet{shapley_1914} cemented Cepheids as pulsating stars (and not, e.g., eclipsing binaries). 
Although the first physical theory of stellar pulsations emerged in a series of 19 publications by \citet{ritter_1879}, this work was largely ignored, and the theory was later independently rediscovered by \citet{eddington_1917_aa}. 
The ensuing century saw further development of linear pulsation theory \citep{baker_1962_aa,Zhevakin_1963_aa,kippenhahn_1969_aa,king_1975_aa,christy_1966_aa,Christy_1968_aa} and eventually nonlinear pulsation theory, including advances in the theory of convection and hydrodynamical modeling of radial pulsations \citep{stellingwerf_1982_aa,Unno_1989_aa, xiong_1997_aa, kollath_1998_aa, yecko_1998_aa, wuchterl_1998_aa, wuchterl_1999_aa, bono_1999_aa, buchler_2001_aa,Kollath_2002_aa, smolec_2008_aa, smolec_2008_ab,geroux_2011_aa,desomma_2024_aa,kovacs_2025_aa,deka_2025_aa}. 

Simulations of stellar evolution show that Cepheid pulsations arise when evolved massive stars enter the shell--hydrogen burning and/or core--Helium burning phase. These stars possess overstable radial modes in their envelopes, which can be driven by the $\kappa$ (opacity valve) mechanism and $\gamma$ (recombination driven) mechanism \citep{becker_1981_aa, cox_1980_aa, xu_2004_aa}. The existence of an instability strip in the Hertzsprung-Russel diagram is a direct result of Cepheid variable stars showing these radial instabilities when evolving through these regions. However, relating the luminosity and period of Cepheids to their evolutionary mass remains a challenging problem to this day \citep{stobie_1969_aa, iben_1972_aa, iben_1975_aa, Moskalik_1992_aa, pietrzynski_2010_aa, neilson_2011_aa,anderson_2014_aa,anderson_2016_aa,miller_2020_aa}. 

Numerical models of Cepheid radial pulsations have so far been limited to static model builders entirely decoupled from stellar evolution calculations, where bespoke nonlinear, time-dependent convection theories are employed to study short timescale pulsation calculations. 
Modern studies of turbulent convection in evolutionary calculations have advanced significantly, but still rely on assumptions of steady-state solutions where turbulence is static \citep{Kupka_2022_aa,ahlborn_2022_aa,braun_2024_aa,deka_2025_ab}.

Recently, \citet{jermyn_2023_aa} incorporated a form of the \citet{kuhfuss_1986_aa} time-dependent convection model into \MESA\ that converges on short and long timescales, allowing smooth transitions from evolutionary phases where convection is static to short thermal and dynamical phases of evolution where the time dependence of convection becomes important. The incorporation of time-dependent turbulent convection into evolutionary calculations allows self-consistent modeling of Cepheid pulsations with stellar evolutionary models. 

The aim of this work is to improve upon the time-dependent convection treatment in \citet{jermyn_2023_aa} (\code{MESA-TDC}) by introducing the eddy-viscous damping term which couples convective motions to pulsation. This work brings the time-dependent convection framework in \citet{jermyn_2023_aa} into closer alignment with the Radial Pulsation module (\code{MESA-RSP}) introduced into \MESA\ in \citet{paxton_2019_aa}.This work does not aim to revise Period-Luminosity, Period-Wesenheit, or related relations, but rather provides a theoretical and numerical framework upon which such investigations could be built in the future.

Section~\ref{s.method} describes the differences between \code{MESA-TDC} and \code{MESA-RSP}, introduces the eddy viscosity into \code{MESA-TDC}, and details the stellar model physics adopted in this work. Section~\ref{s.evol} demonstrates the evolution and pulsation of a stellar model using \code{MESA-TDC} with eddy viscosity, and shows comparisons with \code{MESA-RSP}. Finally, Section~\ref{s.conclusion} summarizes our results. 

\section{Method}\label{s.method}

\subsection{\code{MESA-TDC} and \code{MESA-RSP}}\label{method1}
We begin by summarizing the current formulations for time-dependent convection (TDC) present in \MESA. Both forms of TDC in \MESA\ solve the \citet{kuhfuss_1986_aa} turbulent convection model, a one-equation time-dependent turbulent convection model for quasi-steady state convection. TDC was first introduced into \MESA\ inside the radial stellar pulsation (\code{RSP}) module \citep{paxton_2019_aa}. The \code{RSP} form of TDC was formally introduced by \citet{smolec_2008_aa} and inherited from formulations of the Vienna stellar pulsation hydro code \citep{wuchterl_1998_aa,wuchterl_1999_aa}. Many of these pulsation codes, including \code{RSP}, were built following pioneering work from \citet{stellingwerf_1982_aa}. \code{RSP} shares many similarities to the Florida--Budapest code \citep[see][]{Kollath_2002_aa}. Notably, \code{RSP} includes the effects of negative buoyancy in convectively stable layers, often excluded from other stellar pulsation codes, which has been shown to suppress the appearance of double-mode pulsations \citep{smolec_2008_ab}. \code{RSP} was introduced into MESA with functionality for building static model envelopes for pulsation calculations given global stellar parameters, but does not interface with the stellar evolution portion of the \code{MESA-star} module, though it does adopt the same equations of state and opacities from the \code{eos} and \code{kap} modules. \citet{jermyn_2023_aa} first introduced TDC in the local limit into \code{MESA-star}, which enabled stellar evolution with time-dependent convection that was previously limited to static envelopes.

We will henceforth refer to the \code{RSP} form of this convection model as \code{MESA-RSP}, and the latter general-purpose stellar evolution form of the \code{MESA-RSP} convection model available in \code{MESA-star} as \code{MESA-TDC}. Further details on the differences between these two convection models can be found in \citet{jermyn_2023_aa}.

For this work, we will ignore discussion of the limit of long timescales, where \code{MESA-TDC} makes some simplifying assumptions to reduce to the \citet{cox_1968_aa} form of mixing length theory (MLT).
In this context, MLT is replaced by an equation for the evolution of turbulent kinetic energy $e_{t}$.
Following \citet{kuhfuss_1986_aa}, the time evolution of $e_{t}$ is governed by:
\begin{equation}\label{eq:eturb}
    \frac{D e_{t}}{Dt} = -\alpha_{p_{t}} P_{t} \frac{D \rho^{-1}}{Dt} + \epsilon_q + C - \frac{\partial L_{t}}{\partial m}.
\end{equation}
The first term represents the work done against turbulent pressure, where $\alpha_{p_{t}}$ is a dimensionless free parameter. The energy associated with viscous damping from turbulent eddies is represented by $\epsilon_{q}$, $C$ is the coupling term containing sources and sinks for convection, and $L_{t}$ represents the non-local advection of turbulent flux between neighboring mass shells:
\begin{equation}\label{eq:fturb}
    L_{t} = 4 \pi r^2 (-\alpha \alpha_t \rho h e_{t}^{1/2}) \frac{\partial e_{t}}{\partial r}.
\end{equation}
\code{MESA-RSP} includes $L_{t}$ but \code{MESA-TDC} does not, i.e., Equation~\ref{eq:fturb} is assumed to be zero representing the local-limit solution of convection. 

The eddy-viscous damping term $\epsilon_q$ is:
\begin{equation}\label{eq:eq}
    \epsilon_q = \frac{4}{3} \alpha_m \alpha h e_{t}^{1/2} \left( \frac{d}{dr} \left( \frac{v}{r} \right) \right)^2
\end{equation}
and the coupling term $C = S - D$ where
\begin{equation}\label{eq:s}
    S = \alpha\alpha_{s} e_t^{1/2} \frac{T P Q}{h} \mathcal{Y} \\
\end{equation}
\begin{equation}\label{eq:d}
	D = \alpha_D \frac{e_t^{3/2}}{\alpha h} + \frac{48\sigma \alpha_r}{\alpha^2}\left(\frac{T^3}{\rho^2 c_P \kappa h^2}\right) e_t.
\end{equation}
Here $h=P/(\rho g)$ is the pressure scale height, $\mathcal{Y}~\equiv~\nabla~-~\nabla_{\rm ad}$ is the superadiabatic gradient, and ${Q \equiv \partial \rho^{-1}/\partial T|_P}$ is the thermal expansion coefficient. Equation \ref{eq:d} consists of two parts, the first term representing the dissipation into the turbulent cascade, and the second term representing dissipation from radiative losses. 
The remaining coefficients $\alpha$, $\alpha_{s}$ $\alpha_{d}$, $\alpha_m$, $\alpha_r$, $\alpha_{p_{t}}$, and $\alpha_{t}$ are free parameters.

To correctly incorporate Equation~\ref{eq:eturb} into the stellar structure, one must also include the impact of turbulent pressure $P_{t}$ and eddy viscosity $U_{q}$ in the momentum equations.
The turbulent pressure is evaluated on cells and is given by
\begin{equation}
    P_{t} = \alpha_{p_{t}} \rho e_{\text{t}}
\end{equation}
and the eddy viscosity term is given by
\begin{equation}\label{eq:uq}
    U_q = \frac{1}{\rho r^3} \frac{d}{dr} \left( \frac{4}{3} \alpha_m \rho \alpha h e_{\text{t}}^{1/2} r^3 \frac{d}{dr} \left( \frac{v}{r} \right) \right).
\end{equation}

In \citet{jermyn_2023_aa} the turbulent energy was incorporated into the other equations of stellar structure via heat and momentum transport, but not the total energy equation. In this work, we include the turbulent energy in \MESA's total energy equation. The turbulent pressure $P_{t}$ is included in the momentum equation.

In the luminosity equation we incorporate
\begin{equation}
	L = L_{\rm r}~+~L_{\rm c}~+~L_{\rm t}~,
	\label{eq:L}
\end{equation}
where $L_{\rm r}$ is the radiative luminosity and
\begin{equation}
	L_{\rm c} = 4\pi r^2 \alpha \alpha_{c} \rho c_P T w \mathcal{Y}
	\label{eq:Lconv}
\end{equation}
is the convective luminosity. Here $w \equiv \sqrt{e_t}$ is the turbulent velocity, and the corresponding convective velocity is $v_{c} \equiv \sqrt{\frac{2}{3}}w$. Note that we have chosen $\alpha_{t} = 0$ in Equation~\ref{eq:fturb}, leading $L_t$ to be zero in Equation~\ref{eq:L} in \code{MESA-TDC}, the local limit solution, whereas \code{MESA-RSP} supports the inclusion of non-zero $L_t$. Finally, the luminosity enters the total energy equation, which sets the time evolution of the specific internal energy $e$ in each cell.

\begin{deluxetable}{lll}[!ht]
  \tablecolumns{3}
  \tablewidth{1.0\apjcolwidth}
  \tablecaption{Free parameters of the \code{MESA-TDC} convection model, their base values, and associated \MESA\ controls that multiply the base values. \label{tab:alphas}}
  \tablehead{ \colhead{Parameter \quad $=$} & \colhead{Base Value \quad $\times$} & \colhead{Control Value} }
  \startdata 
    $\alpha$        & $1$    & \code{mixing\_length\_alpha}  \\
    $\alpha_{\rm m}$ & $1$    & \code{TDC\_alpha\_M} \\
    $\alpha_{\rm s}$ & $(1/2)\sqrt{2/3}$ & \code{TDC\_alpha\_S} \\
    $\alpha_{\rm c}$ & $(1/2)\sqrt{2/3}$ & \code{TDC\_alpha\_C} \\
    $\alpha_{\rm d}$ & $(8/3)\sqrt{2/3}$ & \code{TDC\_alpha\_D} \\
    $\alpha_{\rm p_{t}}$ & $2/3$           & \code{TDC\_alpha\_Pt} \\
    $\alpha_{\rm r}$ & $2\sqrt{3}$     & \code{TDC\_alpha\_R} \\
    $\alpha_{\rm t}$ & $1$   & \code{0}
  \enddata
\end{deluxetable}

The values $\alpha_{c} = \frac{1}{2}\sqrt{\frac{2}{3}}$, $\alpha_{s} = \frac{1}{2}\sqrt{\frac{2}{3}}$, $\alpha_{d} = \left(\frac{8}{3}\sqrt{\frac{2}{3}}\right)$, $\alpha_m = 0$, $\alpha_r = 0$, $\alpha_{p_{t}} = 0$, and $\alpha_{t} = 0$ are typically chosen to reproduce the mixing length theory solution in the limit of long timescales. Their base values and associated \code{MESA-TDC} controls are shown in Table~\ref{tab:alphas} which is similar to Table~3 of \citet{paxton_2019_aa} with base values drawn from \citep{kuhfuss_1986_aa,wuchterl_1998_aa}.

\subsection{Changes to \code{MESA-TDC}}

Multiple changes to \code{MESA-TDC} were made to enable this study. In the original \citet{jermyn_2023_aa} formulation of \code{MESA-TDC}, the eddy viscosity terms $\epsilon_{q}$ and $U_{q}$ were not included, as their effects were assumed to be negligible on evolutionary timescales. In this work, we are interested in the dynamical evolution of our stellar models; hence, we have extended \code{MESA-TDC} by including the effects of eddy viscosity in the convection model and in \MESA's momentum equation and energy equations for implicit hydrodynamics.

We adopt the same form as in \cite{smolec_2008_aa}, where Equation~\ref{eq:eq} and ~\ref{eq:uq} can be rewritten in the following Lagrangian form:
\begin{equation}E_q=\frac{4}{3}\frac{1}{\rho}\mu_Q\bigg(\frac{\partial v}{\partial r}-\frac{v}{r}\bigg)^2=4\pi X\frac{\partial (v/r)}{\partial m},\label{eq:Eq2}\end{equation}
\begin{equation}U_q=\frac{1}{\rho r^3}\frac{\partial}{\partial R}\bigg[\frac{4}{3}\mu_Qr^3\bigg(\frac{\partial v}{\partial r}-\frac{v}{r}\bigg)\bigg]=\frac{4\pi}{r}\frac{\partial X}{\partial m},\label{eq:Uq2}\end{equation}
where
\begin{equation}X=\frac{16}{3}\pi\mu_Qr^6\rho\frac{\partial (v/r)}{\partial m},\label{eq:chi}\end{equation}
and
\begin{equation}\mu_Q=\alpha\alpha_m\rho h e_t^{1/2},\label{eq:muq}\end{equation}
is kinetic turbulent viscosity. Implementation of $E_{q}$ and $U_{q}$ into \MESA's discretized equations requires the finite difference stencil to obey strict tridiagonality, as required by \MESA's BCYCLIC block tridiagonal solver \citep{hirshman_2010_aa,paxton_2013_aa}. This restricts the stencil for $E_{q}$ and $U_{q}$ to coupling at most three cells, without introducing solver instability. We are able to form $E_{q}$ (viscous heating) and $U_{q}$ (viscous acceleration) on faces as needed by \code{MESA-TDC} and \MESA's default momentum equation for implicit hydrodynamics (\code{v\_flag}). For \MESA's energy equation we mass average $E_{q}$ from faces to cell centers, instead of rebuilding $E_{q}$ with a cell centered stencil, although both approaches are numerically indistinguishable. 

\MESA{} also offers a Godunov type hydrodynamic scheme (\code{u\_flag}), which adopts the Hartan-Lax-van Leer-Contact hydrodynamic solver \citep[HLLC,][]{1994_Toro_ams, paxton_2018_aa}. At the moment, $U_{q}$ as written in Equation~\ref{eq:Uq2} is difficult to introduce into the HLLC solver without widening the stencil to four cells and/or introducing solver instability.We address this by adopting a staggered stencil for radii, when forming $\partial(v/r)$ in Equation \ref{eq:chi}. This approach provides numerical stability in HLLC at the cost of reduced spatial accuracy from second to first order in $U_{q}$. The eddy viscosity can now be included in \code{MESA-TDC} (for \code{v\_flag} and \code{u\_flag}) with the new \code{`TDC\_alpha\_M'} control similar to the \code{`RSP\_alfam'} control available in \code{MESA-RSP}.

In the original \citet{jermyn_2023_aa} formulation of \code{MESA-TDC}, the turbulent energy was not included in the total energy equation, rather implicitly coupled through the total luminosity $L$. In this work, we allow turbulent energy from \code{MESA-TDC} to feedback into the total energy equation, although the impact is minor and the tradeoff is numerical stability on long timescales. This can be toggled with the control
\code{`TDC\_include\_eturb\_in\_energy\_equation = .true.'}. In this work, we find that including turbulent energy in the total energy equation increases the energy error incurred per timestep by approximately one order of magnitude per timestep versus models without this added energy (\code{log\_rel\_error\_in\_energy\_conservation}~$\sim -9$ versus $\sim -10$ during pulsation calculations). Disparities in energy conservation could be larger in stellar models not explored in this work, and could depend sensitively on mesh and time resolution. It is important to note that that including turbulent energy in the total energy equation becomes prohibitive for convergence in stellar evolutionary models, hence our evolutionary calculations do not include this feedback. It is likely that most stellar evolution calculations, with their large implicit timesteps, are prohibited from including this additional, small source of energy due to its numerical instability.

\subsection{Physical and numerical differences between \code{MESA-TDC} and \code{MESA-RSP}}\label{s.differences}

Both \code{MESA-RSP} and \code{MESA-TDC} differ in their numerical implementations. While both \code{MESA-RSP} and \code{MESA-TDC} adopt the \citet{kuhfuss_1986_aa} convection model, \code{MESA-RSP} solves the energy equation on cell centers, whereas \code{MESA-TDC} solves this turbulent energy equation locally and analytically on each cell face. 

\code{MESA-TDC} as implemented in \citet{jermyn_2023_aa} includes a correction to the convective efficiency, allowing it to asymptotically reproduce the \citet{cox_1968_aa} formulation of MLT in the limit of long timescales. In this work, we do not include this correction. The correction can be removed by toggling the new control \code{`include\_mlt\_corr\_to\_TDC = .false.'}. \code{MESA-TDC} then operates similar to \code{MESA-RSP} by directly solving Equations~\ref{eq:eturb}, \ref{eq:L}, and \ref{eq:Lconv} implicitly with the other equations for stellar structure and hydrodynamics. 

One major difference comes from the numerical solving technique. \code{MESA-RSP} adopts $e_{t}$ as a solver variable and uses thermodynamic relations to form $\mathcal{Y}$ using finite differences on the grid. In contrast, \code{MESA-TDC} inverts the luminosity equation and uses Equation~\ref{eq:L} and \ref{eq:Lconv} to solve for the $\mathcal{Y}$ needed to carry the flux. The inverted luminosity equation \code{MESA-TDC} approach is numerically more stable than \code{MESA-RSP} because $L$ is very sensitive to small changes in $\mathcal{Y}$, but $\mathcal{Y}$ is relatively insensitive to small changes in $L$. This difference is extremely important because the accuracy of $\mathcal{Y}$ in \code{MESA-RSP} is set by the grid scale resolution since $\mathcal{Y}$ is computed from finite differences on the Lagrangian grid. In contrast, \code{MESA-TDC} can directly solve for the $L_{c}$ necessary to carry the flux through the cell face interfaces between two zones and so does not suffer from under or overestimating the convective flux in unresolved regions. 

Notably, whether using \code{MESA-RSP} or \code{MESA-TDC} the hydrodynamics equations are time centered to overcome the numerical dissipation resulting from backward Euler integration. As numerical residuals are driven to zero, time centering offers intrinsic energy conservation, preserving modes that develop in the solution. Time centering can also preserve spurious numerical modes induced by imperfect operations such as remeshing. These modes can induce ringing across the mesh and are numerically unstable; hence, without a substantive adaptive mesh refinement scheme, the mesh is frozen after initial remeshing and remains fixed for all calculations done in \code{MESA-RSP} and \code{MESA-TDC}. While time centering is not always necessary to overcome numerical dissipation in the envelopes of extremely luminous stars with large mode growth rates, it remains necessary to resolve the growth of pulsations explored in this work. Future improvements such as the implementation of a second-order stable implicit time integration scheme into \code{MESA} could potentially offer the benefit of both damping high frequency noise while still resolving the growth of physical nonlinear radial pulsations. 

\subsection{Stellar model physics}\label{s.input_physics}

This section describes the relevant stellar physics in our evolutionary models. These stellar evolution models serve as inputs into our stellar pulsation modeling setup detailed in Section~\ref{s.pulsation_setup}. All of our calculations are conducted in a development version of MESA. Our changes to \code{MESA-TDC} will be available in the next \code{MESA} release. The \MESA\ files and plotting scripts to reproduce our results are available at \dataset[http://doi.org/10.5281/zenodo.17945924]{http://doi.org/10.5281/zenodo.17945924}.

Our stellar models are evolved from the pre-main sequence to the third instability strip crossing, and then the timestep is decreased to conduct our pulsation experiments. We adopt an initial mass of 6 \Msun\, helium abundance ${Y=0.267}$ and metallicity ${Z=0.003}$, characteristic of Cepheids in the Large Magellanic Cloud. Metal abundances are scaled using the \citet{asplund_2009_aa} solar abundance mixture. 
We adopt radiative opacities from \citet{iglesias_1993_aa,iglesias_1996_aa,ferguson_2005_aa,poutanen_2017_aa}, conductive opacities from \citet{cassisi_2007_aa, blouin_2020_aa}, and the equation of state as a blend between \citealt{saumon_1995_aa, timmes_2000_ab, rogers_2002_aa, irwin_2004_aa, potekhin_2010_aa, jermyn_2021_aa, bauer_2023_aa}. 
The evolutionary models adopt an Eddington-gray atmosphere as an outer boundary condition, with a relaxed surface optical depth boundary condition of $\tau = 10^{-3}$. 
Relaxation of the atmosphere to low optical depth is not a necessary process, but we make this choice as RSP does this by default and our goal in this work is to compare and contrast stellar models drawn from \code{MESA-star} as closely as possible. 
When studying the dynamical behavior of our models, we do not include an additional atmosphere and instead adopt a zero pressure and a free-streaming luminosity boundary condition similar to \code{MESA-RSP}, which can be adopted with the control \code{use\_RSP\_L\_eqn\_outer\_BC = .true.}. 
However, our stellar pulsation models using \code{MESA-TDC} are not limited to this free-streaming luminosity outer boundary condition, and fully support the inclusion of atmospheres and other hydrodynamic boundary conditions available in \code{MESA-star}. Future investigations may probe stellar pulsation models with varying outer boundary conditions. 
Our stellar evolution models employ the ``Reimers'' wind mass loss scheme \citep{reimers_1977_aa} with an efficiency factor of 0.5. Mass loss is deliberately included to remove material with inhomogeneous composition arising in the extended, optically-thin layers of the atmosphere. As we discuss later, our stellar pulsation models in \code{MESA-TDC} retain inhomogeneities in chemical composition generated from stellar evolution.
We do not include mass loss in our subsequent stellar pulsation models; however, our models do support the inclusion of mass loss, which may allow for future explorations of the pulsation and mass-loss coupling. 

For convection, all models in this work adopt the parameters from Model~A in \citet{paxton_2019_aa}, as shown in Table~\ref{tab:adopted_alphas}.

\begin{deluxetable}{lc}[!ht]
  \tablecolumns{2}
  \tablewidth{1.0\apjcolwidth}
  \tablecaption{Free parameters of the \code{MESA-TDC} convection model adopted in this work. \label{tab:adopted_alphas}}
  \tablehead{ \colhead{Control Value} & \colhead{This Work} }
  \startdata 
    \code{mixing\_length\_alpha} & 1.5 \\
    \code{TDC\_alpha\_M} & 0.25 \\
    \code{TDC\_alpha\_S} & 1 \\
    \code{TDC\_alpha\_C} & 1 \\
    \code{TDC\_alpha\_D} & 1 \\
    \code{TDC\_alpha\_Pt} & 0 \\
    \code{TDC\_alpha\_R} & 0 
  \enddata
\end{deluxetable}

The adoption of $\alpha_{m} = 0.25$ as a fiducial value is consistent with \code{MESA-RSP} and is somewhat motivated by agreement with observed Cepheid amplitudes \citep[see][]{deka_2025_ab}. This choice was also made to retain continuity with earlier \code{MESA-RSP} calibrations; however, the correct choice for $\alpha_{m}$ remains largely empirical, and its value is typically calibrated on a per-star basis so as to produce the desired amplitude consistent with a given observation. Since stellar evolution calculations are carried out in the local limit (i.e., ${\alpha_{t}=0}$), convective core overshooting is included via the \citep{herwig_2000_aa} exponential diffusive overshooting scheme with ${f_{\rm ov}=0.01}$ and ${f_{0,{\rm ov}}=0.005}$. Our stellar pulsation calculations neglect convective overshooting. 

During stellar evolution, we adopt nuclear reaction rates from a combination of NACRE \citep{angulo_1999_aa} and JINA REACLIB \citep{Cyburt_2010_ab}. Reaction rate screening corrections are from \citet{chugunov_2007_aa}, a dynamic screening method which includes a physical parametrization for the intermediate screening regime and reduces to the weak \citep{dewitt_1973_aa, graboske_1973_aa} and strong \citep{alastuey_1978_aa,itoh_1979_aa} screening limits at small and large values of the plasma coupling parameter. Weak reaction rates are based, in order of precedence, on  \citet{langanke_2000_aa}, \citet{oda_1994_aa}, and \citet{fuller_1985_aa}.

\subsection{Stellar pulsation setup}\label{s.pulsation_setup}

This section describes the setup of our pulsation calculations. We carry out three types of stellar pulsation calculations. The first involves freezing the mesh of the full stellar model and retaining the nuclear burning core; the second involves generating an envelope model by excising the core and remeshing the envelope to match the grid used by \code{MESA-RSP}; and the third involves generating a static envelope model with identical properties from \code{MESA-RSP} and evolving it in both \code{MESA-TDC} and \code{MESA-RSP} so direct comparisons can be made. 

Envelope models are generated from the stellar evolution calculation by removing the core below a specified interior temperature of $T = 2\times 10^{6}$~K, using the \MESA\ star\_job control \code{remove\_center\_by\_temperature~=~2d6}. The internal boundary condition for the envelope is set by the interior mass, and the luminosity, which is held fixed. The stellar envelope is then remeshed following an identical scheme to that discussed in \citet{smolec_2008_aa,paxton_2019_aa}. The envelope is remeshed such that surface zones contain equal mass until a specified anchor point temperature is reached: we adopt $T_{anchor} = 11,000$~K, corresponding to just below the zones for hydrogen ionization. Above this temperature anchor, zones geometrically increase in mass inward until the interior boundary condition at $T = 2\times 10^{6}$~K is reached. Following the resolution recommendations from \citet{paxton_2019_aa}, the remeshed envelopes adopt the highest resolution explored to obtain reasonable convergence, which is a total of 600 zones: 240 zones between the surface and $T_{anchor} = 11,000$~K, and 360 zones between the anchor and the interior boundary condition. We offer users controls for remeshing the envelope. These remeshing options are provided via the following new inlist controls:\\
\\
\code{TDC\_hydro\_nz = 600}\\
\code{TDC\_hydro\_nz\_outer = 240}\\
\code{TDC\_hydro\_T\_anchor = 11d3}\\

\begin{figure} 
    \centering
    \includegraphics[width=3.15in]{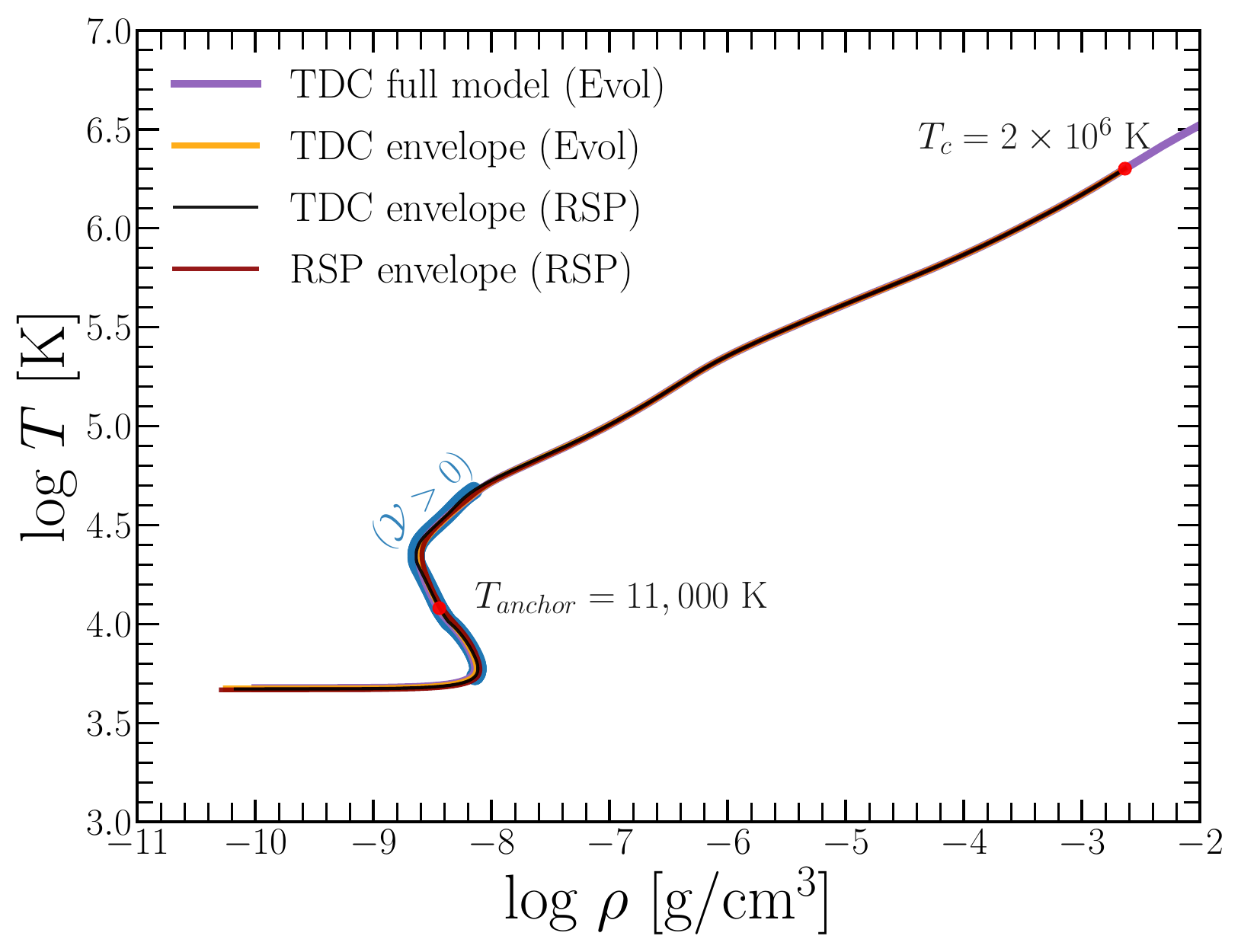}
    \caption{Initial stellar model logarithmic $\rho-T$ profiles for the \Mzams\,=\,6\Msun\ \code{\MESA-star} full stellar evolution model (purple) and the remeshed stellar evolution evolution model envelope (yellow), both labeled ``Evol.'' The envelopes created in the \code{MESA-RSP} model builder (blue and red) are labeled ``RSP''. The temperature anchor and interior boundary conditions for the envelope models are marked, and the convection zone where $\mathcal{Y} > 0$ is highlighted in blue. All four models display an almost identical structure.}
    \label{fig:initial_model_profiles}
\end{figure}

The logarithmic $\rho-T$ profiles of all four stellar models are displayed in Figure~\ref{fig:initial_model_profiles}, demonstrating nearly identical structures. 

For models generated from stellar evolution, we utilize version 8.0 of the \code{GYRE} stellar oscillation software instrument \citep{townsend_2013_aa,townsend_2018_aa, goldstein_2020_aa,sun_2023_aa}. 
We identify unstable nonadiabatic ${l = 0}$ (radial) modes by their positive growth rates. 
If the fundamental mode eigenfunction has a positive growth rate, we initialize the stellar model in the fundamental mode with a 5~km/s velocity kick, similar to what is done in \code{MESA-RSP}. 
This procedure is carried out using \code{run\_star\_extras} hooks. 
It should be noted that \code{GYRE} currently adopts the frozen-in flux approximation for convection, so the linear nonadiabatic analysis (LNA) provided by this tool is not equivalent to \code{MESA-RSP}'s LNA, which includes perturbations to the convection flux in its analysis. 
This difference is only important for identifying if modes are unstable, and the largest differences between both LNA tools likely occur near the red edge of the instability strip. 
The limit cycle amplitude and behavior reached via nonlinear integration of these hydrodynamical envelope models is unaffected by the adopted linear analysis tool.  Mode selection can be affected by the type of initial perturbation delivered to the model, however we are only concerned with fundamental mode pulsation in this work and we do not directly explore mode selection. In either case, we initialize our stellar pulsation models with some kinetic energy in the fundamental mode, as is done in \code{MESA-RSP}.

For static envelope models generated in \code{MESA-RSP}, we use the LNA to initialize our stellar model in the fundamental mode with a 5~km/s velocity kick. The \code{MESA-TDC} model loaded from \code{MESA-RSP} constructed model is initialized after the kick, so the initial perturbation is retained.

During all pulsation calculations, for reasons described in Section \ref{s.differences}, the mesh is held fixed during all pulsation calculations. All models adopt the Von Neumann Richtmeyer form of artificial viscosity with $C_{q} = 4$ and $\alpha = 0.1$ as reasonable choices which do not result in dissipation except for the strongest shocks, \citep[impact of viscosity on Cepheids has been studied by][]{takeuti_1998_aa}.
If time centering of the equations was not needed, as in for example the model of larger $L/M$ envelopes, then the mesh could be in principle unfrozen.

\section{Evolution and Pulsation in \code{MESA-star}}\label{s.evol}

\begin{figure} 
    \centering
    \includegraphics[width=3.15in]{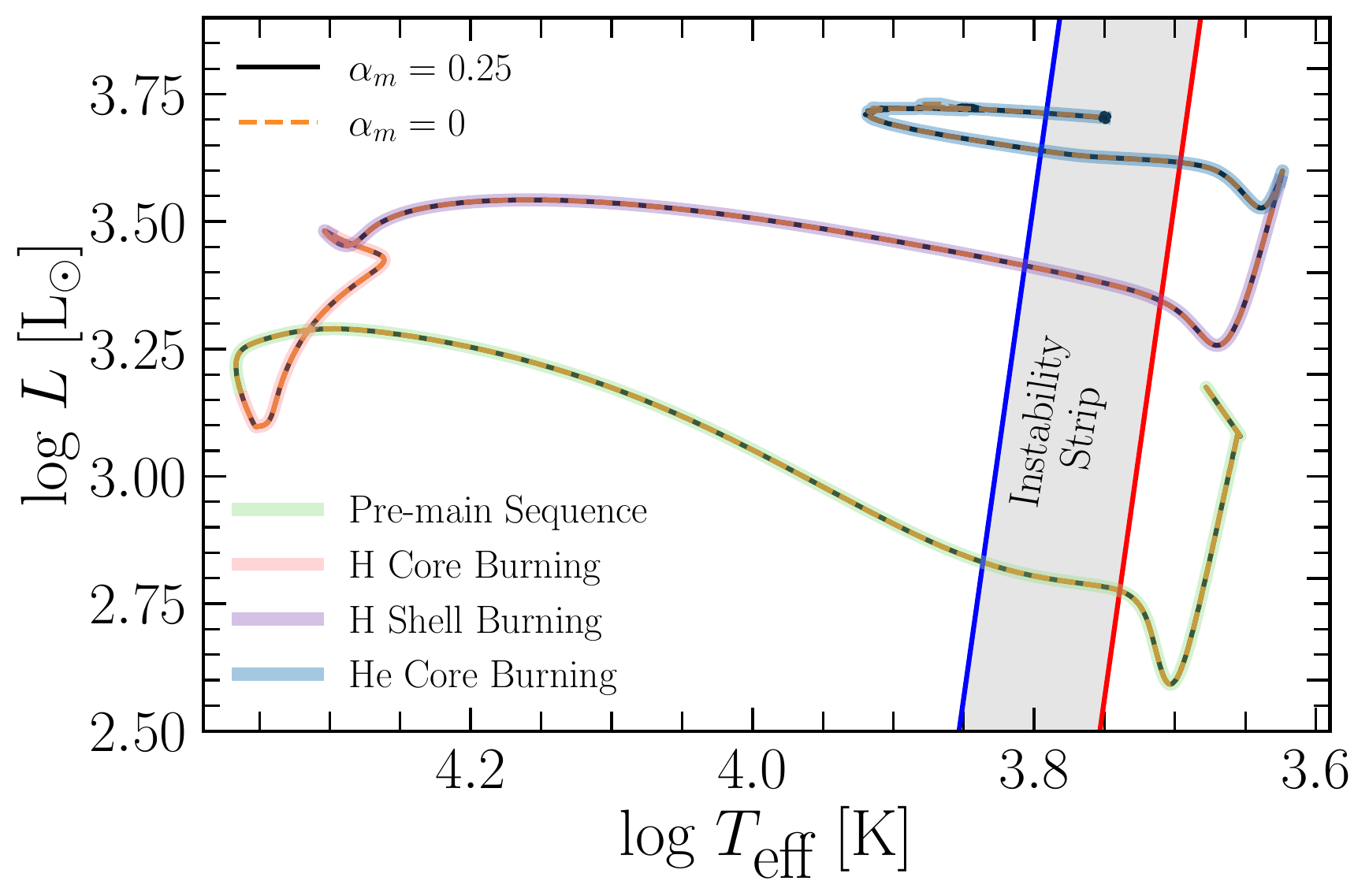}
    \includegraphics[width=3.15in]{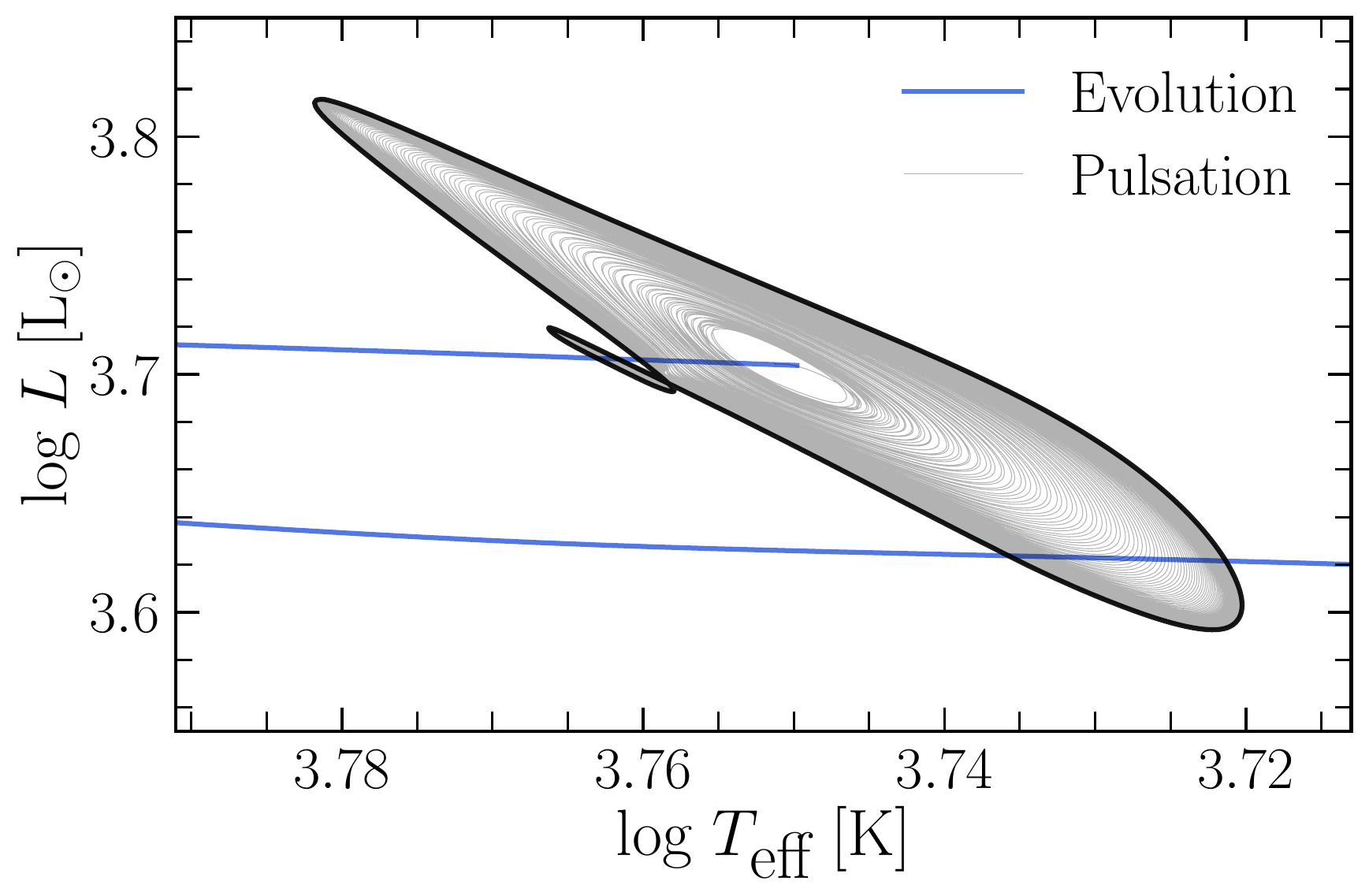}
    \includegraphics[width=3.15in]{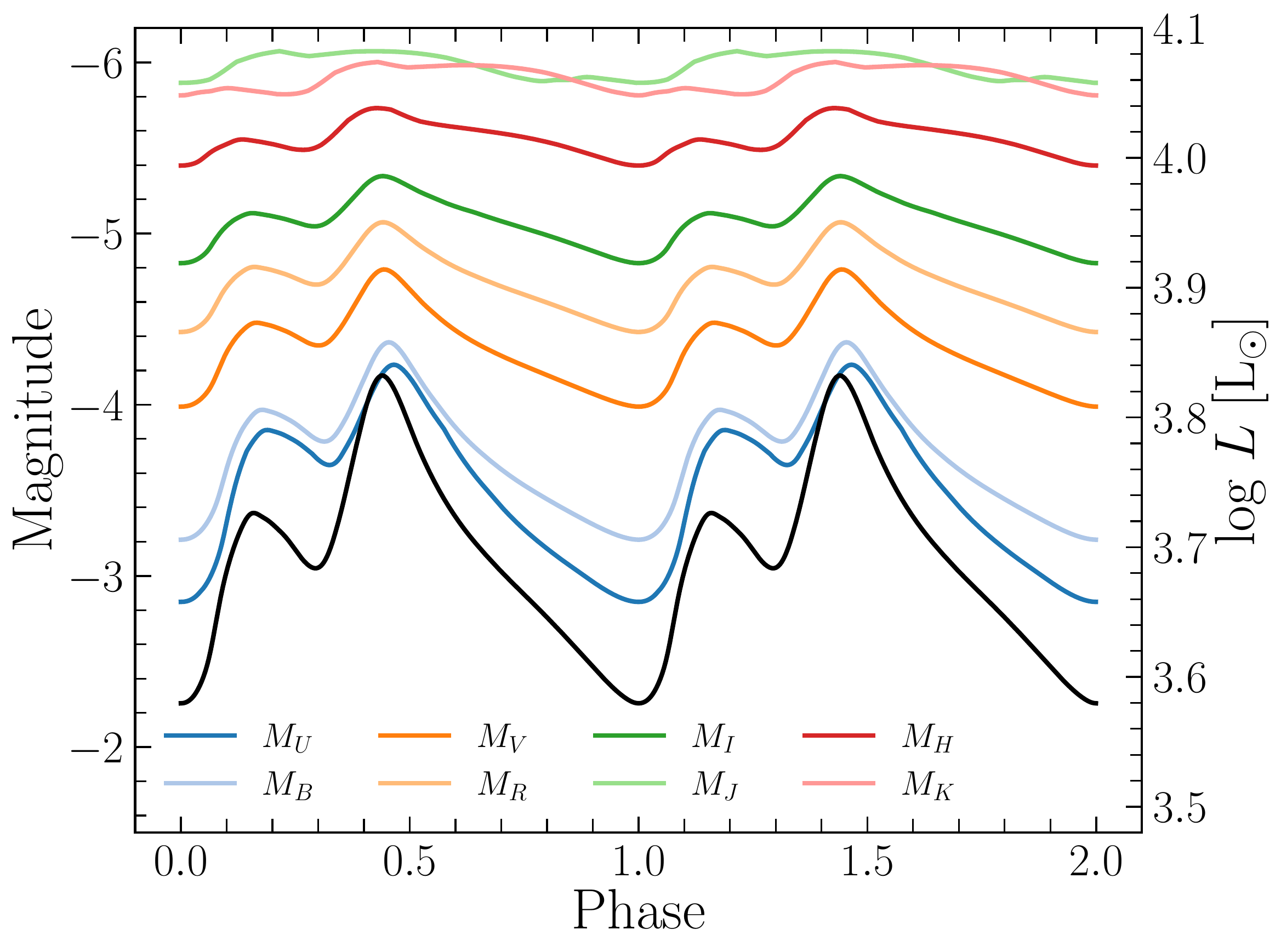}
    \caption{(Top panel) HRD of the \Mzams\,=\,6\Msun\ \code{\MESA-star} evolution model using \code{TDC} with and without the eddy-viscous damping \code{`TDC\_alpha\_M = 0.25d0, 0d0'}, color coded by phase of evolution. An approximate location for the instability strip is shown as a gray region between blue and red edges. (Middle panel) A close-up of the Cepheid radial pulsation growth (thin grey lines) to finite amplitude pulsations (thick black line) in the HRD overlaid on the evolution track. (Bottom panel) Light curves of the Cepheid pulsations in various photometric bands (left y-axis) and log $L/$\Lsun\ plotted in black (right y-axis), both shown as a function of pulsation phase.}
    \label{fig:hrd_evol}
\end{figure}

In this section, we evolve two $\Mzams\ = 6~\Msun$ hydrodynamical stellar models from the pre-main sequence until the core-He burning phase. The models are evolved until reaching $\log \Teff/\textrm{K} = 3.75$ during the third crossing of the classical Cepheid instability strip. The models are then paused and their structure is used to evolve a pulsation perturbation until a stable full amplitude pulsation is reached. 

One stellar model includes the effects of the physical eddy-viscous damping in the turbulent energy and momentum equations ($U_{q}, E_{q}$) with $\alpha_{m} = 0.25$ (\code{`TDC\_alpha\_M = 0.25d0'}), and the other does not. The evolution of both stellar models is illustrated in the Hertzsprung-Russell Diagram (HRD) on the top panel of Figure~\ref{fig:hrd_evol} with approximate locations of the blue and red edges of the instability strip shown. As assumed in \citet{jermyn_2023_aa}, we confirm here that the eddy viscosity has a negligible effect on the stellar evolution of a hydrodynamical model. For most of the evolution, the radial velocities in these stellar models remain $\lesssim 10^{-6}~c_{s}$, where $c_{s}$ is the local sound speed. This is expected because most stellar evolution is conducted in or near hydrostatic equilibrium. Low velocities during these long evolutionary phases result in the eddy viscosity being a negligible contribution to the convection model except during dynamical phases of evolution. Nonetheless, we illustrate the strong agreement between both stellar models with and without eddy viscosity to suggest its numerical robustness, and hence we suggest to include it in all hydrodynamical calculations as it is a physical component of the convection model.

Once the evolution model reaches $\log \Teff/\textrm{K} = 3.75$, the calculation is paused with global stellar envelope parameters of ${M = 5.9089}$~\Msun, ${\Teff = 5620}$~K, log${L/L_{\odot} = 3.703}$. The chemical abundance mass fractions ${X = 0.267395}$, ${Z = 0.003032}$ are slightly altered from their initial values due to the first convective dredge up which occurs on the Red Giant Branch (RGB). Whether the core is excised and/or the envelope is remeshed, chemical profiles acquired from stellar evolution are preserved when conducting calculations in \code{MESA-TDC}. For the scenario of Cepheids, the envelope is very homogeneous so composition gradients are not present or explored in this work. However, our \MESA\ setup allows for the exploration of pulsation models with inhomogeneous composition profiles generated from stellar evolution, enabling future studies of mixing processes in the context of radial stellar pulsations.

The middle panel of Figure~\ref{fig:hrd_evol} displays a close-up of the evolution through the instability strip in the HRD. Overlayed on top, is the growth of the pulsation perturbation to the stellar pulsation model after a 5 km/s fundamental mode eigenfunction kick is delivered. The \code{MESA-TDC} stellar pulsation model grows until reaching saturation with $\Delta$ log $L/$\Lsun~$\simeq 0.3$  resulting in a final nonlinear fundamental mode period $P = 11.886$~days. We assume the stellar pulsation to be saturated once the fractional growth in radial kinetic energy growth over each pulsation cycle drop below $10^{-6}$. At this point the cycle to cycle variation in period of the stellar pulsation is constant to within $10^{-3}$ seconds.

In the bottom panel of Figure~\ref{fig:hrd_evol}, we illustrate the light curve in different photometric bands using \code{MESA}'s color module, with colors from \citet{lejeune_1998_aa} over two full phases of stellar pulsation. The $I, J, K$ and $H$ bands are the brightest, however, the visible $V$ and $U$ bands more closely resemble the bolometric light curve (i.e., log~$L/$\Lsun). The local maximum before maximum light that occurs during the contraction phase is a known feature in Cepheid light curves resulting from a harmonic overtone, a 2:1 resonance with the second radial overtone \citep{simon_1976_aa}. Its position with respect to maximum light depends sensitively on the pulsation period. Cepheids with this bump feature are referred to as ``Bump Cepheids,'' and changes in the pulsation period of these types of stars is reflected in the relative location of this feature with respect to maximum light. The dependence of this feature on stellar pulsation period is referred to as the ``Hertzsprung progression" \citep{hertzsprung_1926_aa,simon_1981_aa}.

\subsection{Eddy viscosity}\label{s.eddy}

\begin{figure} 
    \centering
    \includegraphics[width=3.35in]{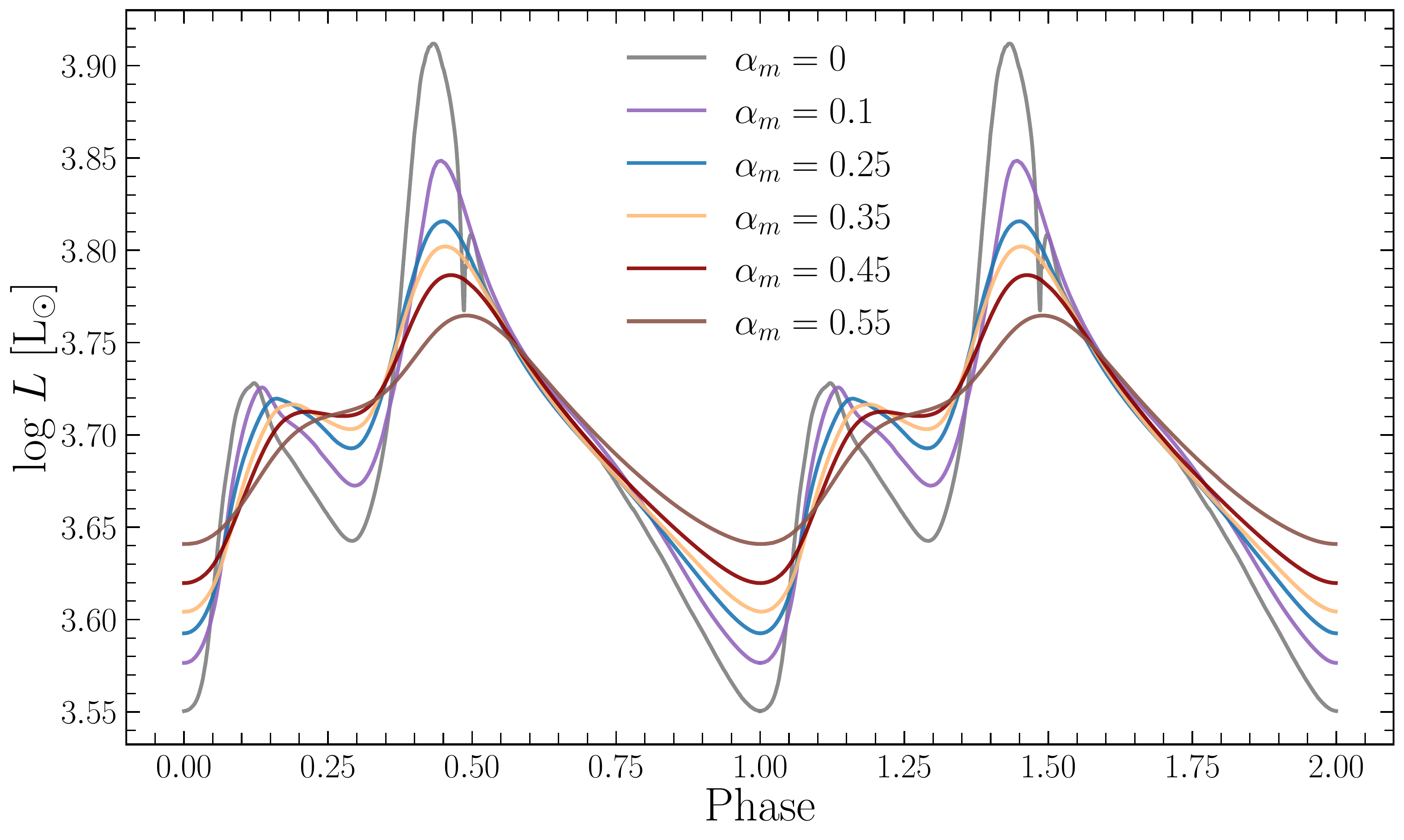}
    \caption{Logarithmic luminosity light curves for the \Mzams\,=\,6\Msun\ \code{\MESA-TDC} model initialized from stellar evolution with different values of eddy viscosity (\code{`TDC\_alpha\_M = 0d0, 0.1d0, 0.25d0, 0.35d0, 0.45d0, 0.55d0'}).}
    \label{fig:alfam_evol}
\end{figure}

In this section, we verify that our implementation of the eddy viscosity (see Equations \ref{eq:eq} and \ref{eq:uq}) alters the amplitude and shape of our light curves. Using the same initial evolution model as in Section~\ref{s.evol}, we evolve three separate pulsation models, each with the different value of $\alpha_{m}$, which controls the strength of eddy-viscous dissipation. In Figure~\ref{fig:alfam_evol}, we illustrate the light curves against stellar pulsation phase over two cycles for an identical initial model with $\alpha_{m} = 0, 0.1, 0.25, 0.35, 0.45$~and~$0.55$. Our results are consistent with Figure~6 in \citet{paxton_2019_aa}, wherein larger values of $\alpha_{m}$ result in enhanced eddy-viscous dissipation which limits the cycle-to-cycle amplitude and smooths out the bump feature while having little effect on the overall pulsation period The eddy viscosity captures the coupling between convection and pulsation; hence, without it, pulsations models have difficulty settling on a single mode. For this reason,  models with $\alpha_{m} <  0.25$, take more than twice as long to reach finite amplitude in the fundamental mode. Likewise, the lack of viscous damping in models with $\alpha_{m} <  0.25$ leads to larger, less physical amplitudes. On the other hand, models with larger eddy viscous dissipation also take significantly longer to achieve saturation than the $\alpha_{m} =  0.25$ model, as the kinetic energy growth rate per cycle also decreases with increasing $\alpha_{m}$. This makes sense physically, as eddy viscous dissipation appears in the momentum equation as a sink for momentum, decreasing the total average work done per cycle. This corresponds to a lower fraction of kinetic energy growth rates per cycle. The $\alpha_{m} =  0.55$ model takes roughly four times as many cycles  as the $\alpha_{m} =  0.25$ model to reach saturation (600 versus 150). We tested a $\alpha_{m} =  0.65$ model, however after $> 2000$ cycles, saturation was not reached, and the amplitudes remained nearly an order of magnitude lower than the $\alpha_{m} =  0.55$ model. Nonetheless, the final periods for all five models are very similar, remaining within 0.025 days of one another, and decreasing with increasing $\alpha_{m}$. This corresponds to $\delta P/P \lesssim 3\times10^{-3}$. We caution however that this need not always be the case, as very nonadiabatic stellar models, such as those of red super giants and/or asymptotic giant branch stars, could display a period dependence on $\alpha_{m}$ not captured in this work.

\subsection{Comparison with \MESA-RSP }\label{s.compare}

\begin{figure}
\centering
\includegraphics[width=3.38in]{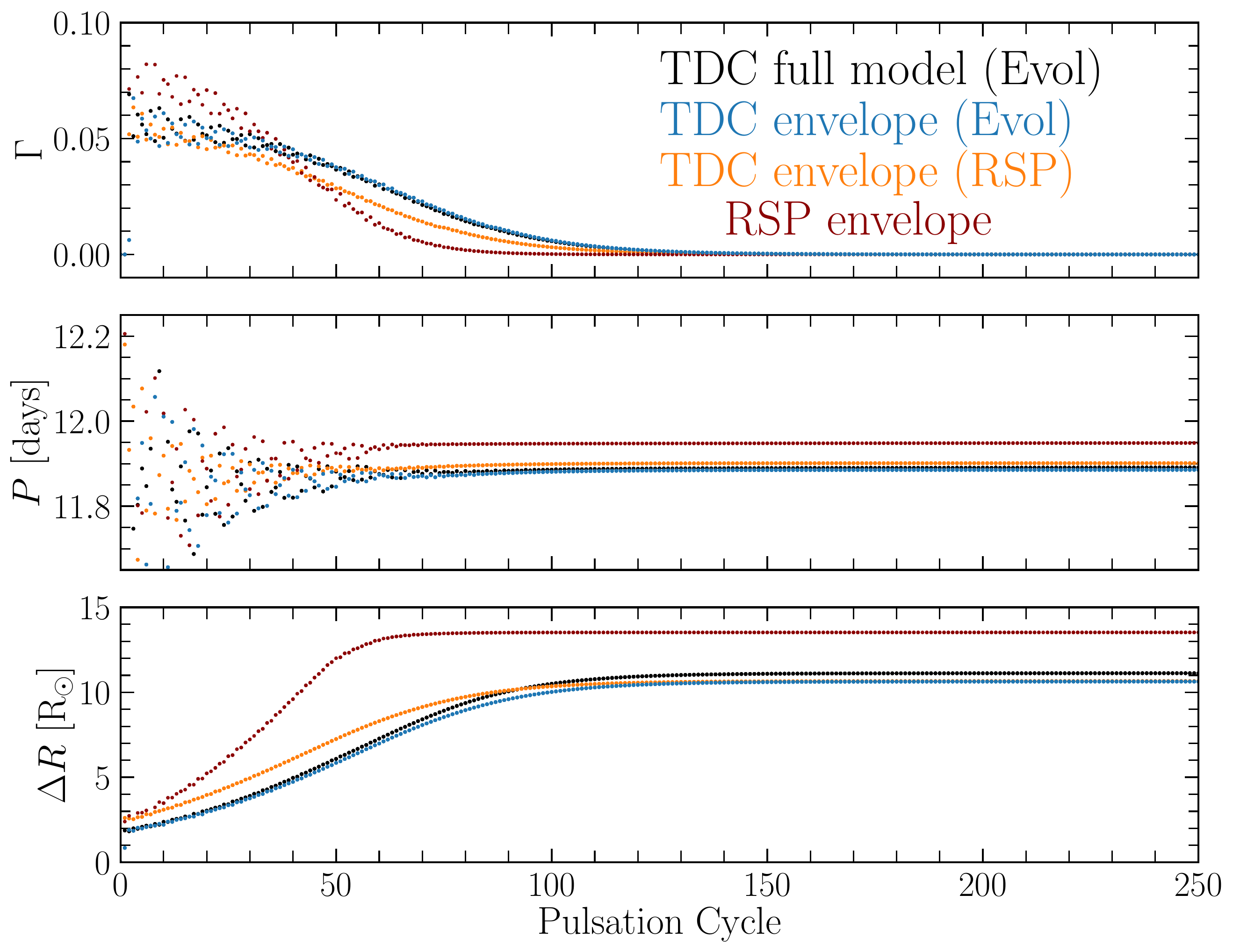} \\
\caption{ Global cycle-to-cycle properties of four models: the \code{MESA-TDC} model initialized from stellar evolution and both \code{MESA-TDC} and \code{MESA-RSP} pulsation model initialized from a \code{MESA-RSP} static envelope. Fractional kinetic energy growth rates $\Gamma$, periods $P$, and radial variation $\Delta R$ are plotted versus pulsation cycle in the top, middle, and bottom panels, respectively. Model properties are provided in Table~\ref{tab:table3}. All models adopt $\alpha_{m} = 0.25$.}
\label{fig:compare_growth}
\end{figure}

We now focus on comparing our \Mzams~=~6~\Msun\ Cepheid in \code{MESA-TDC} with an identical model in \code{MESA-RSP}, assuming the same physics and boundary conditions. We continue using the stellar evolution models from Section $\ref{s.evol}$ as a reference. We take the stellar model properties from this model and initialize a \code{MESA-RSP} with the same initial parameters. After building the initial model in \code{MESA-RSP} with these properties, we kick the model in the fundamental radial eigenmode with 5~km/s surface velocity, and then load this initial model back into \code{MESA-TDC} and evolve it along side the \code{MESA-RSP} model to compare how the two identical initial models evolve differently in \code{TDC} versus \code{RSP}. The final properties of all four models are displayed in Table~\ref{tab:table3}. We find that models run with \code{MESA-TDC} take roughly twice as long to be computed. This is likely due to \code{MESA-TDC} utilizing the \code{MESA-star} infrastructure and therefore carrying more overhead as well as many additional equations for the chemical composition not evolved in \code{MESA-RSP}. While there is a computation trade-off to adopting \code{MESA-TDC}, it comes with added physical capabilities not limited to composition gradients, mass loss or gain, binary evolution, gravitational settling, radiative levitation, nuclear reactions, and rotation, reinforcing why this trade-off is justified. It is also important to emphasize that a variety of stars evolve through the HRD out of thermal equilibrium. Hence, drawing pulsation models from stellar evolutionary tracks can be more accurate than employing static models, which typically generate fully relaxed initial models in thermodynamic equilibrium. The full stellar model pulsation calculation has $\sim 1800$ stellar model zones, and therefore takes an additional factor of three times longer to compute than any other model studied in this work. This is the penalty of including the stellar model core.

\begin{deluxetable*}{lcccccccc}[!htb]
  \tablenum{3}
  \tablecolumns{9}
  \tablecaption{
   Stellar pulsation model properties\label{tab:table3}}
  \tablehead{
    \colhead{} &
    \colhead{$P$} &
    \colhead{$\Delta M_{bol}$} &
    \colhead{$\Delta R$ (\Rsun)} &
    \colhead{$\phi_{21}$} &
    \colhead{$\phi_{31}$} &
    \colhead{$R_{21}$} &
    \colhead{$R_{31}$} &
    }
\startdata
        \code{TDC (Full Evol model)} & 11.894 d & 0.5546  & 11.137 & 1.183 & 6.202 & 0.124 & 0.381 \\
        \code{TDC (Evol envelope)} & 11.886 d & 0.5578  & 10.635 & 1.065 & 0.018 & 0.106 & 0.338\\
        \code{TDC (RSP envelope)}  & 11.901 d & 0.5637  & 10.659 & 1.002 & 0.029 & 0.102 & 0.331 \\ 
        \code{RSP (RSP envelope)}  & 11.949 d &  0.5924 &  13.523 & 1.551 & 0.767 & 0.242 & 0.517 \\
  \enddata
\end{deluxetable*} 

\begin{figure*}
\centering
\includegraphics[width=6.38in]{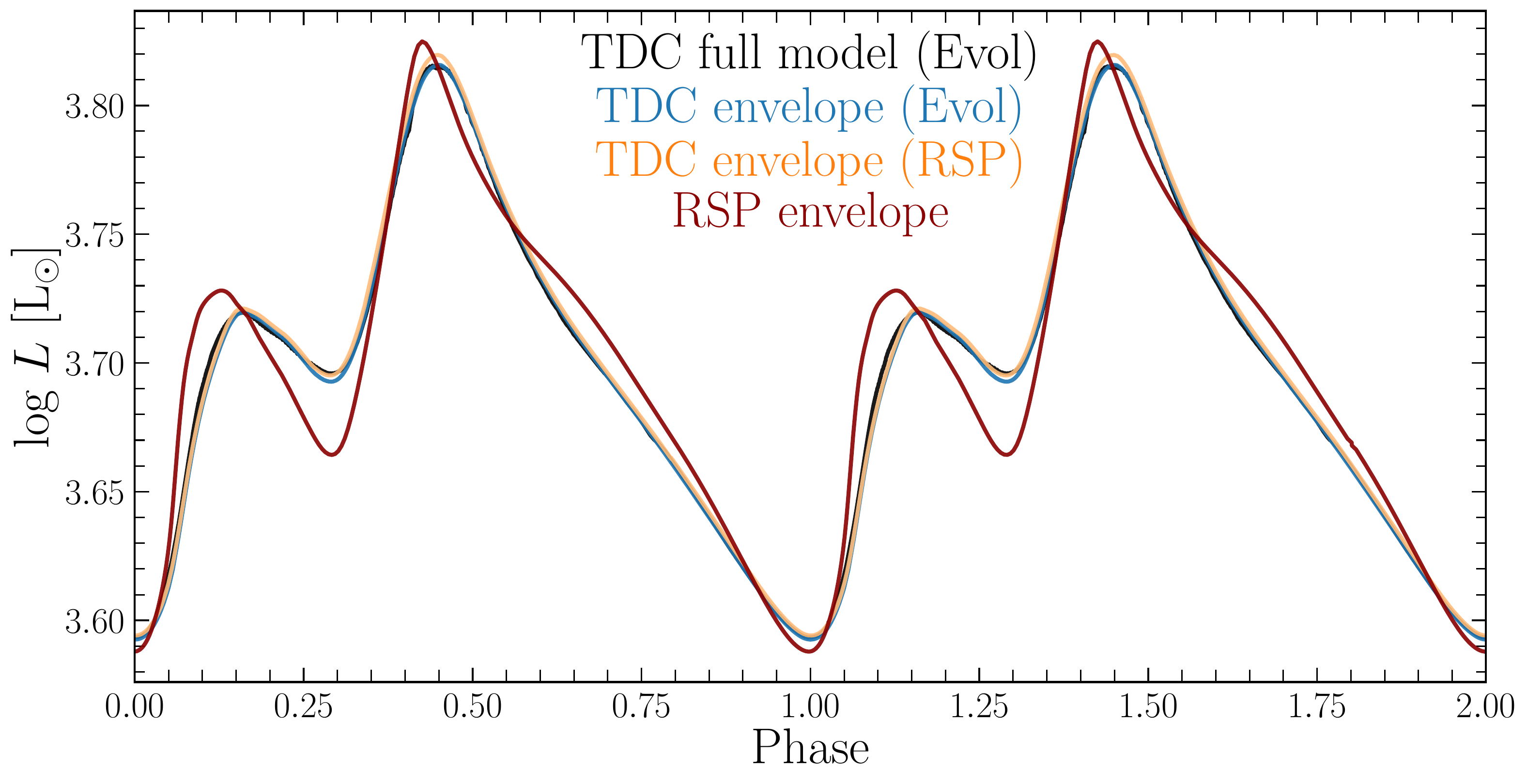} \\
\includegraphics[width=6.38in]{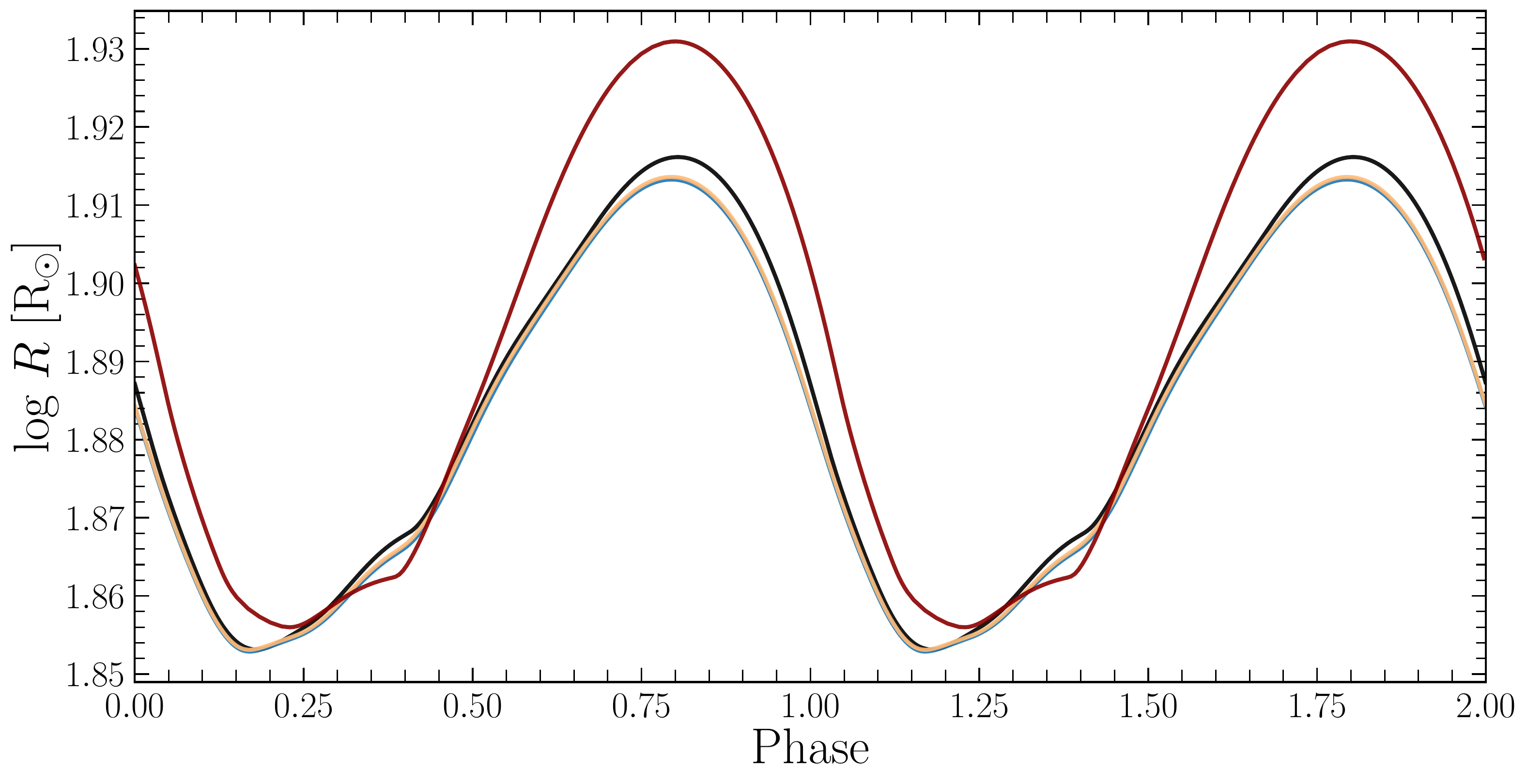} 
\caption{Logarithmic luminosity (top panel) and radius (bottom panel) versus pulsation phase for all four models in Table~\ref{tab:table3}. All models adopt $\alpha_{m} = 0.25$. }
\label{fig:compare_lightcurves}
\end{figure*}

The global properties of all four models during each pulsation cycle are shown in Figure~\ref{fig:compare_growth}, and luminosity and radius versus pulsation phase are shown in Figure~\ref{fig:compare_lightcurves}. The four models have pulsation periods that differ by less than 0.07~d, corresponding to a fractional change of $\sim0.5$\% (see Table~\ref{tab:table3}), and the morphologies of their light curves are also similar. We also compare the I-band Fourier parameters $\phi_{31}, \phi_{31}, R_{21}$ and $R_{31}$, computed by fitting a Fourier series to the I-band lightcurve via
\begin{align}
    I(t)
    &=
    I_0
    +
    \sum_k
    A_k
    \sin(k\omega t+\phi_k)
\end{align}
from which we derive
\begin{align}
    \phi_{k1} &= \phi_k - k\phi_1\\
    R_{k1} &= A_k / A_1.
\end{align}

Notably, the \code{MESA-TDC} model loaded from stellar evolution appears more similar to the \code{MESA-TDC} model loaded from \code{MESA-RSP} than the \code{MESA-RSP} model itself. The full \code{MESA-TDC} stellar model deviates from the \code{MESA-TDC} envelope models mostly in peak radius, likely because the frozen mesh in the full stellar model possess a distinct and more fine grid spacing, especially near the surface. The bump feature in the light curve in the top panel of Figure~\ref{fig:compare_lightcurves} is considerably more pronounced in the \code{MESA-RSP} model as well. In the bottom panel where the logarithmic radius is displayed, the \code{MESA-RSP} shows a considerably larger radius than all of the \code{MESA-TDC} models.

\begin{figure}
\centering
\includegraphics[width=3.38in]{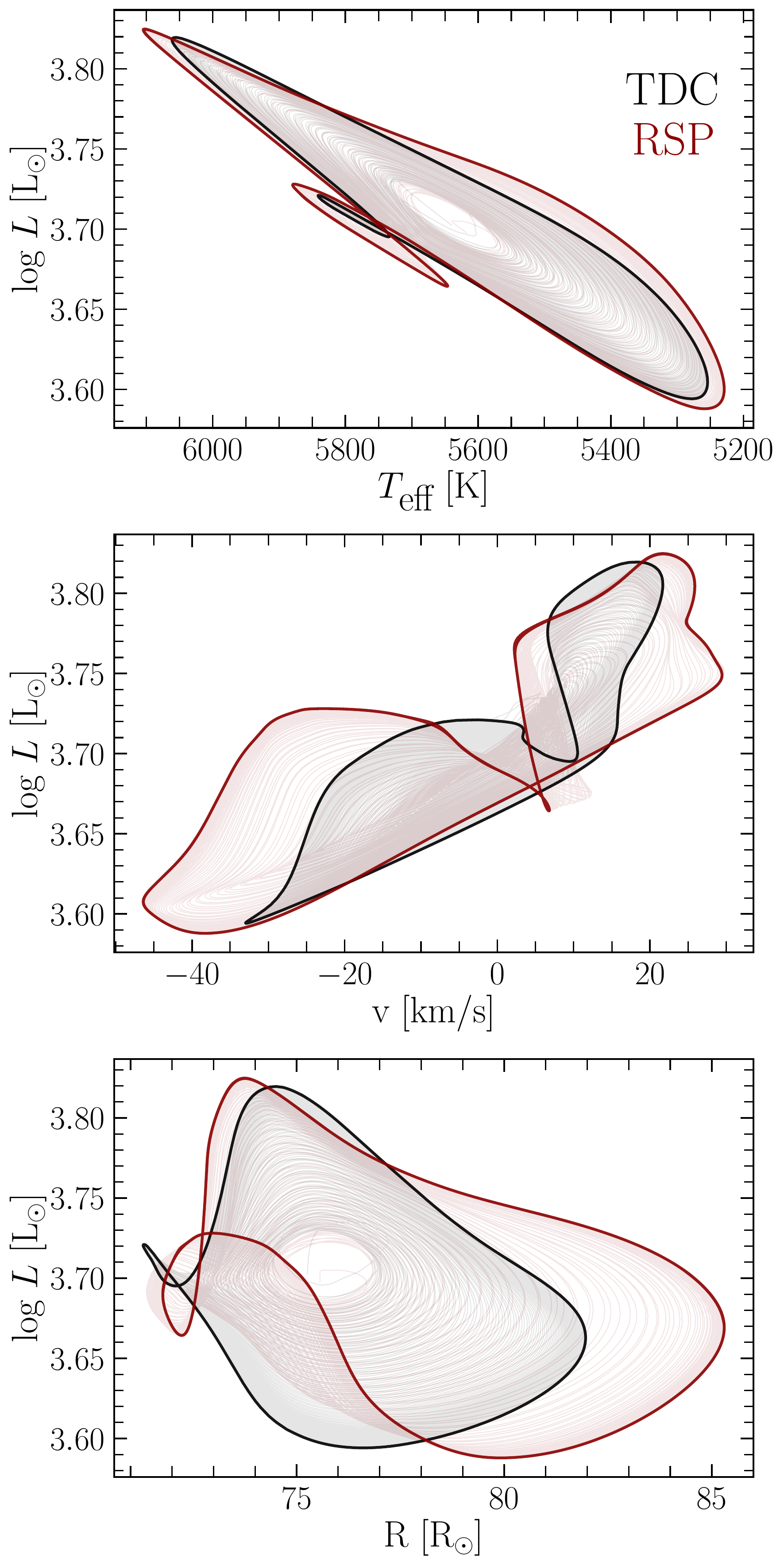} 
\caption{Total logarithmic luminosity $L$  against the effective temperature $T_{\rm eff}$, velocity $v$, and radius $R$ for the \code{MESA-TDC} and \code{MESA-RSP} loaded from an identical initial model, generated in \code{MESA-RSP}. Each Panel shows the growth from the initial perturbation into finite amplitude pulsations with thin lines and the final pulsation cycle with thick, dark lines. Model properties are provided in Table~\ref{tab:table3}. Both models adopt $\alpha_{m} = 0.25$.}
\label{fig:compare_pulse_panel}
\end{figure}

To compare \code{MESA-TDC} and \code{MESA-RSP} more directly, we will focus on the two models loaded from an identical initial RSP envelope. First, we examine the global properties over each pulsation cycle in Figure~\ref{fig:compare_pulse_panel}. Each panel shows the growth from the initial perturbation into finite amplitude pulsations. In the top panel, we display the overlaid HRDs for both models, which look remarkably similar in morphology. The minor differences in the HRD panel are more easily visible in the corresponding log $L/\texttt{\Lsun}$ versus velocity (middle panel) and log $L/\texttt{\Lsun}$ versus radius (bottom panel) plots. The radii and velocities shown in this Figure are from the last model zone, and are not photospheric quantities. These panels illustrate that the \code{MESA-RSP} model consistently gives larger absolute surface velocities, especially during the contraction phase near minimum light when the surface moves with $> 10$ km/s larger inward velocities. During the expansion phase, both the top and bottom panels show the \code{MESA-RSP} model achieving significantly larger maximum radii than the TDC model, followed by a more violent relaxation indicated by the larger peak velocities. Despite these differences, the HRD for both models look very similar. The systematically larger radius and surface velocity excursions found in \code{MESA-RSP} are not numerical artifacts, rather \code{MESA-RSP} systematically overestimates pulsation amplitudes in the Cepheid models explored in this work. We believe these differences result from differences in internal numerical dissipation and the adoption of Equation \ref{eq:L} in \code{MESA-RSP} in place of a temperature gradient equation in \code{MESA-TDC}.

\begin{figure*}
\centering
\includegraphics[width=3.38in]{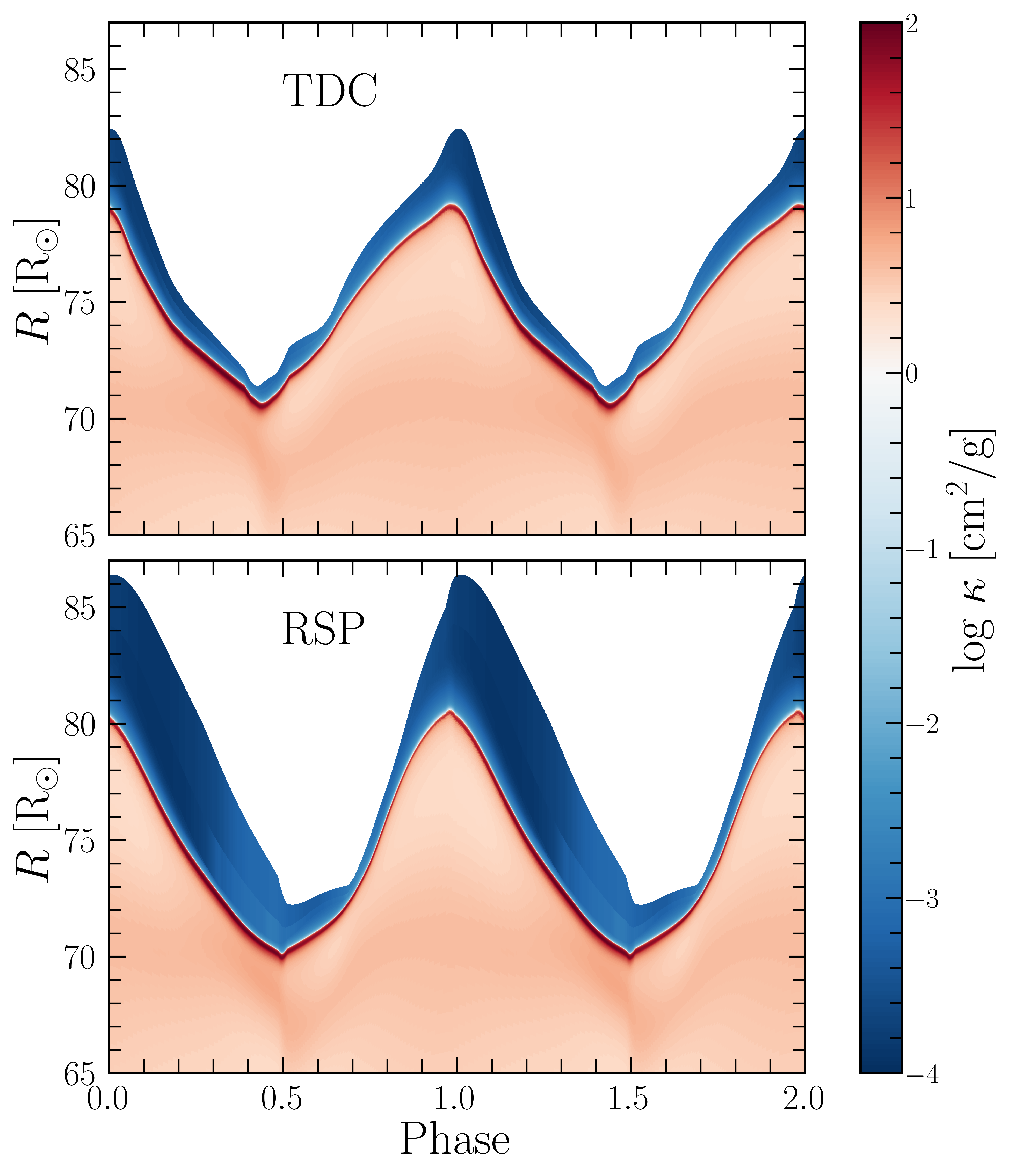} 
\includegraphics[width=3.38in]{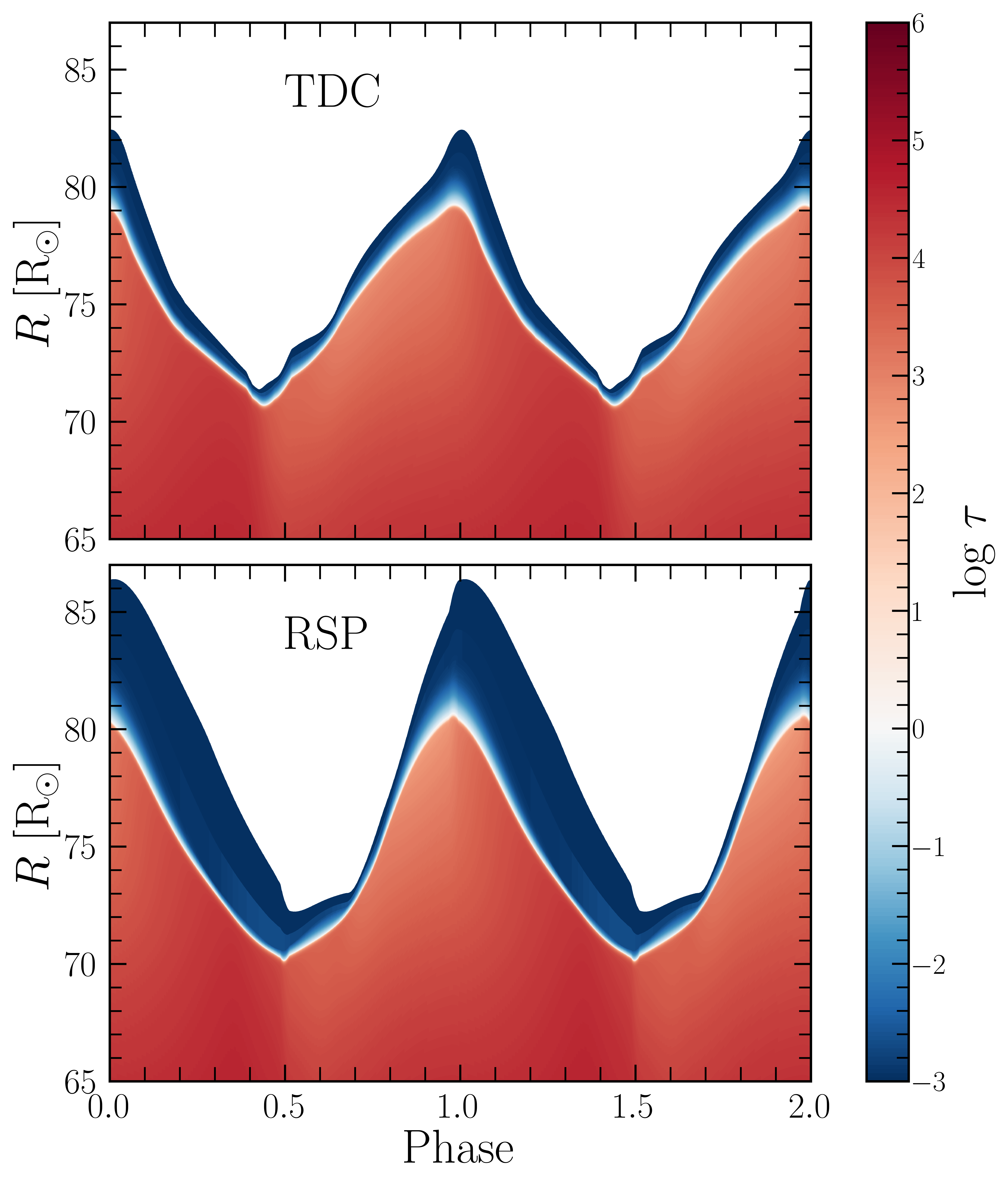} 
\caption{Profiles of the stellar structure across the pulsation cycle for the \code{MESA-TDC} and \code{MESA-RSP} loaded from an identical initial model generated in \code{MESA-RSP}. All panels illustrate the radius versus pulsation phase, with left panels colored by opacity and right panels colored by optical depth. Model properties are provided in Table~\ref{tab:table3}. Both models adopt $\alpha_{m} = 0.25$.}
\label{fig:compare_profiles}
\end{figure*}

\begin{figure*}
\centering
\includegraphics[width=3.38in]{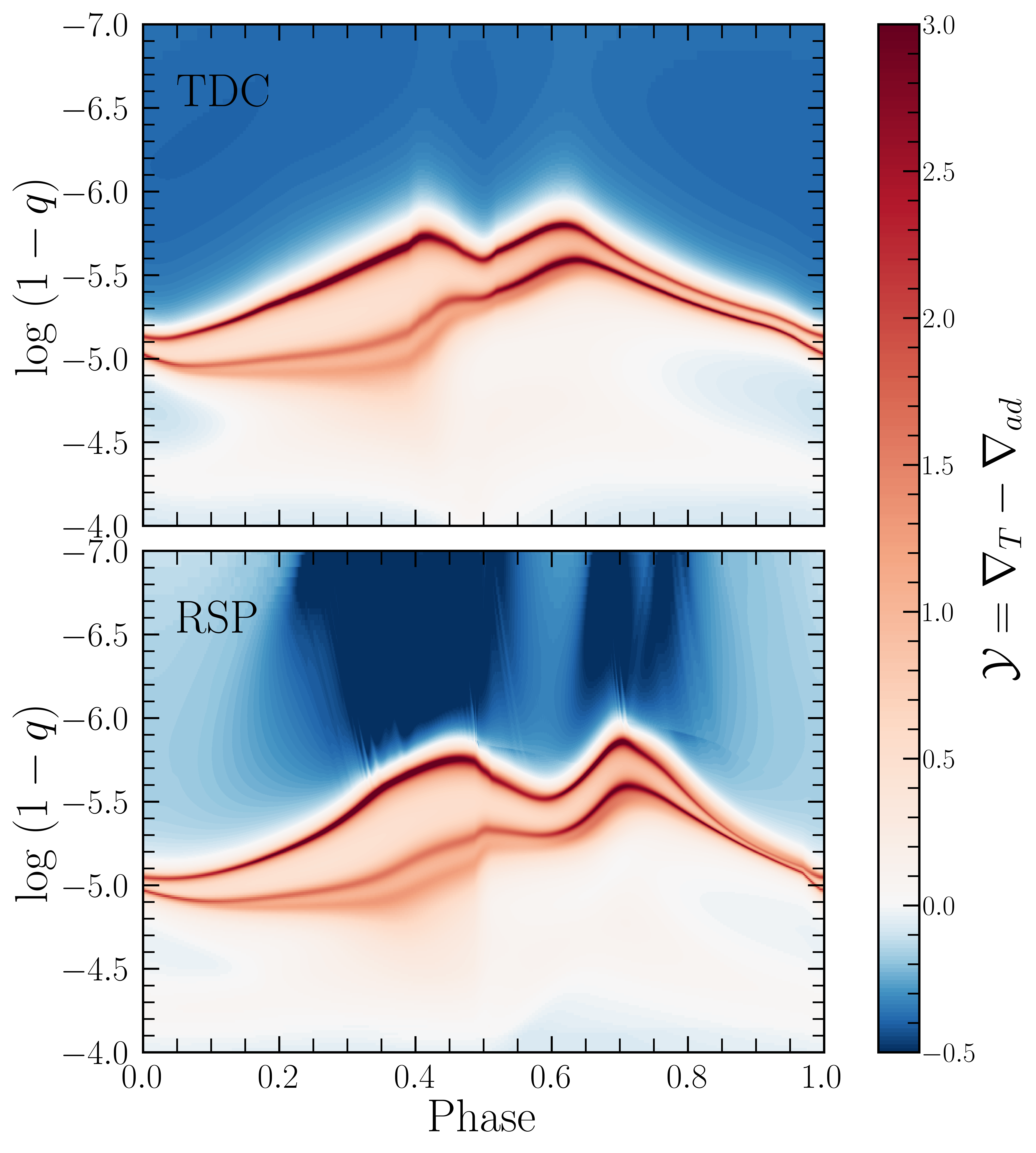} 
\includegraphics[width=3.38in]{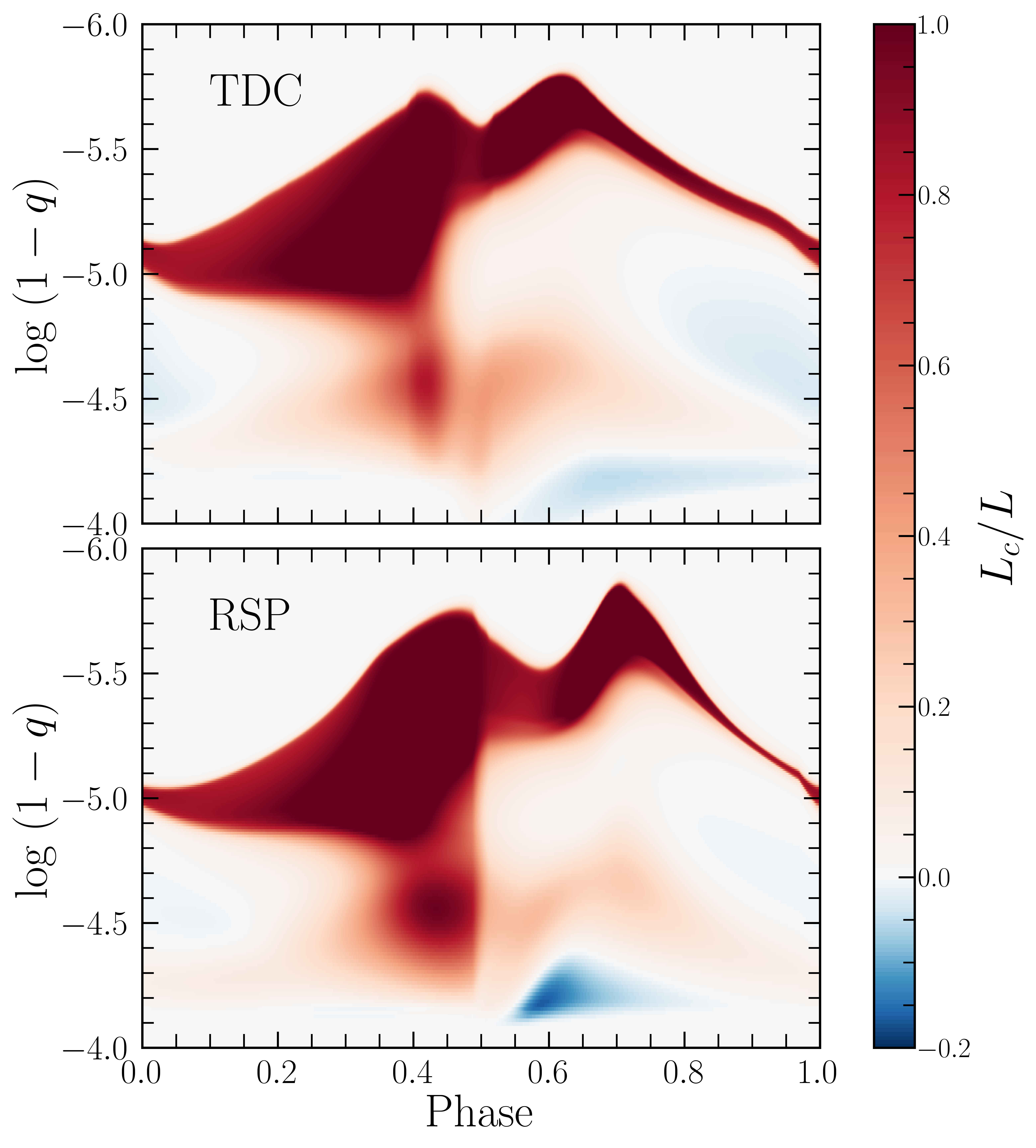} 
\caption{Profiles of the stellar structure across the pulsation cycle for the \code{MESA-TDC} and \code{MESA-RSP} loaded from an identical initial model generated in \code{MESA-RSP}.  All panels illustrate the transformed logarithmic fractional mass coordinate log $1-q$ (where $q = M/M_{\star}$) versus pulsation phase. Left panels are colored by superadiabatic gradient $\mathcal{Y}$ and right panels are colored by the fractional convective luminosity $L_{c}/L$. Dark red regions generally mark the boundary of the Hydrogen ionization bump in the left column, and convective regions in right column. Model properties are provided in Table~\ref{tab:table3}. Both models adopt $\alpha_{m} = 0.25$.}
\label{fig:compare_profiles2}
\end{figure*}

We further examine the structure of these stellar models versus pulsation phase in Figure~\ref{fig:compare_profiles}. This Figure shows both models contain photospheres that occur over a narrow range in radius, and are effectively located near the hydrogen ionization opacity bump. While the atmosphere ($\tau<2/3$) of the \code{MESA-RSP} model is significantly more extended than the \code{MESA-TDC} model, the photospheric radii remain within 1 R$_{\odot}$ of one another at maximum and minimum.

While difficult to investigate exact differences, we conjecture that the primary difference between both models lies in the way the superadiabaticity $\mathcal{Y} = \nabla - \nabla_{\rm ad}$ is determined (see Section~\ref{s.differences}). In \code{MESA-RSP}, there is no temperature gradient equation; instead, Equation~\ref{eq:L} is enforced, where $L_{c}$ is formed from Equation~\ref{eq:Lconv} and is set by $\mathcal{Y}$ which is calculated directly from the grid using finite differences to determine the actual superadiabaticity from EOS quantities. The analytic solution for $\mathcal{Y}$ produced by \code{MESA-TDC} is formed self-consistently on faces. This difference means that when cell faces become radiative in \code{MESA-TDC}, the radiative temperature gradient can be directly enforced on the face, whereas in \code{MESA-RSP} the finite difference formulation for $\mathcal{Y}$ will return similar subadiabatic values, but will depend more sensitively on the grid resolution. While not directly explored in this work, some differences between \code{MESA-TDC} and \code{MESA-RSP} could also arise due to differences in internal numerical dissipation.  

We explore the differences in $\mathcal{Y}$ and fractional convective luminosity $L_{c}/L$ versus pulsation phase and fractional mass coordinate $q = M/M_{\star}$ in Figure~\ref{fig:compare_profiles2}. In the left column $\mathcal{Y}$ appears to be similar in morphology although slight differences in $\mathcal{Y}$ are visible at different depths in the stellar envelope. Near the hydrogen ionization bump, at $T \sim 10,000$~K and between log~$\tau \sim (0, 1)$, the \code{MESA-RSP} model appears to slightly over/underestimate the $\mathcal{Y}$ as compared to \code{MESA-TDC}. The $\mathcal{Y}$ from \code{MESA-RSP} contains substantially more numerical artifacts visible in the dark blue hue where $\mathcal{Y} < 0$. Nonetheless, the bulk fractional convective luminosity profiles $L_{c}/L$ are considerably similar in morphology across both models, with the \code{MESA-RSP} model having a slightly larger maximum and minimum $L_{c}/L$ than the \code{MESA-TDC} model. From this figure we can clearly see that there are differences between \code{MESA-TDC} and \code{MESA-RSP}, and that \code{MESA-TDC} appears more well behaved in optically thin regions. However, we caution from advising the use of \code{MESA-TDC} over \code{MESA-RSP} in general, as the \code{MESA-RSP} formulation could still offer other numerical advantages, aside from computational speed, not explored or discussed in this work. The main conclusion is that these two frameworks are very similar, albeit slightly different in verifiable ways. However, for many Classical Cepheid applications based on static envelopes, \code{MESA-RSP} remains entirely adequate and \code{MESA-TDC} should be viewed as a complementary tool enabling new classes of problems rather than an incremental improvement in standard Cepheid modeling.

\section{Conclusion}\label{s.conclusion}
In an effort to more closely integrate time-dependent convection physics in \MESA, we have introduced an additional eddy-viscous dissipation term previously omitted from the convection model used in \code{MESA-star} introduced in \citet{jermyn_2023_aa}, called \code{MESA-TDC}, but included in the separate \code{MESA-RSP} module \citep{paxton_2019_aa}. We verify its inclusion evolving a \Mzams = 6 \Msun\ stellar model and self-consistently initializing it into fundamental nonlinear radial pulsations. We demonstrate the effectiveness of this eddy-viscous dissipation parameter in controlling the amplitude and shape of radial pulsation light curves. We directly compare an identical initial model between \code{MESA-TDC} and \code{MESA-RSP} to showcase the differences that arise from different numerical methods for solving the same \citet{kuhfuss_1986_aa} 1-equation time-dependent convection model. We conjecture that the primary difference between both of these methods lies in the way the superadiabaticity $\mathcal{Y} = \nabla - \nabla_{\rm ad}$ is determined. In \code{MESA-RSP} $\mathcal{Y}$ is calculated directly from the grid using finite differences to determine the actual superadiabaticity from EOS quantities. In \code{MESA-TDC}, $\mathcal{Y}$ is determined as the solution to a residual luminosity equation which analytically relates the total luminosity to the convective luminosity $L_{c}$. Likewise, we find that $\mathcal{Y}$ as predicted by \code{MESA-RSP} appears to contain more numerical artifacts than the $\mathcal{Y}$ produced by \code{MESA-TDC}. 

With the incorporation of eddy viscosity into \code{MESA-TDC}, we have demonstrated the ability for \code{MESA-star} to produce self-consistently excited radial pulsations directly from stellar evolution calculations. \code{MESA-star} also provides other options not available in \code{MESA-RSP}, including but not limited to various inner and atmospheric boundary conditions, mass loss during the pulsation, binary evolution, rotational deformation, and hooks for implementing other physics options. Optionally including perturbations to the convective flux in the linear nonadiabatic analysis provided by \code{GYRE} is a future project. Likewise, the inclusion of a static model builder in \code{MESA-star} and the nonlocal turbulent flux into \code{MESA-TDC} are also subjects for future work.

For many applications \code{MESA-RSP} remains a working and well tested standard for conducting stellar pulsation calculations. \code{MESA-TDC} is a complementary tool optimized for physical completeness rather than computational efficiency. Choosing one approach over the other will depend on the specific scientific question being explored. Since both \code{MESA-RSP} and \code{MESA-TDC} produce similar light curves, the practice of taking global parameters from stellar evolution models and using them in \code{MESA-RSP} remains a well warranted and reasonable practice. However, \code{MESA-TDC} is likely the better tool to investigate phenomena such as:

\begin{itemize}
    \item Pulsations of stars for which chemical inhomogeneities that arise during evolution are important. The best example are Blue Large Amplitude Pulsators (BLAPs), which are believed to be driven due to the iron bump that arises through radiative levitation. With \code{MESA-RSP}, this behavior has been mimicked by adopting a larger $Z$ in the whole envelope. However, with \code{MESA-TDC}, a model structure directly inherited from evolutionary calculations could potentially generate more reliable BLAP models. Other examples include peculiar double-mode RR Lyrae stars, which are difficult to explain assuming they are indeed ordinary RR Lyrae stars \citep{prudil_2017_aa}, and stars for which binary interactions such as mass transfer have altered their thermal structure or chemical profile. The ability to introduce chemical inhomogeneities in the envelope and the resulting impact on pulsations is a way to verify some ideas about these and other stars. 
    
    \item The direct application of evolutionary-pulsation coupling, such as the modeling of mode switching. There are few classical pulsators that have been observed to mode switch, and for which the switching process occurs very quickly, see \citet{soszynski_2014_aa,soszynski_2014_ab}. This phenomenon has only been studied in non-evolving envelopes as in \code{MESA-RSP} and via amplitude equation (AE) models \citep[e.g.][]{szabo_2004_aa}. However, with access to evolutionary calculations run with \code{MESA-TDC}, one might be lucky enough to find mode switching, or can try to model the phenomenon and compare the amplitude evolution with observations. Then by fitting the AEs, reliable values of parameters that enter the AEs (cross and self-saturation parameters) can help us learn something about mode selection. However studies of such phenomenon  might require less simplified convection models, that can reliably model double-mode pulsations.
    
    \item Pulsation--mass-loss coupling is likely another important phenomenon which can be studied with \code{MESA-TDC}, whether to explore the impact of smooth mass loss on long timescales, or pulsation-induced mass loss in short cycle bursts.

    \item While not explored in this work, \code{MESA-TDC} offers the ability to model more massive large $L/M$ pulsators which \code{MESA-RSP} has difficulty modeling or is completely unable to model. This includes many potential avenues for studying the pulsations of $\delta$~Scuti, $\alpha$~Cygni, S~Doradus, ultra long-period Cepheid, and red supergiant stars. This framework offers potential avenues for expanding upon recent studies of massive star pulsations \citep{bronner_2025_aa,laplace_2025_aa,suzuki_2025_aa,gautschy_2025_aa}.

\end{itemize}

Our results presented in this work are based on a single stellar mass and metallicity, and conclusions found in this work might not generalize to other masses and metallicities. A broader parameter-space exploration of mass, metallicity, and instability-strip crossings is deferred to future work.

\section*{Acknowledgments}\label{s.ack}
EF thanks Bill Paxton and Adam Jermyn for their previous \MESA\ development work and written discussions, without which this work would not be possible. EF also thanks Teresa Braun for her helpful conversations. The authors thank the anonymous referee for helpful and constructive feedback. This research is supported by the Yale Center for Astronomy and Astrophysics Prize Fellowship (EF). RS acknowledges support from the National Science Center, Poland, Sonata Bis grant 2018/30/E/ST9/00598. This research made extensive use of the SAO/NASA Astrophysics Data System (ADS).


\software{
\MESA\ \citep[][\url{https://docs.mesastar.org}]{paxton_2011_aa,paxton_2013_aa,paxton_2015_aa,paxton_2018_aa,paxton_2019_aa,jermyn_2023_aa},
\code{MESASDK} 20250801 \citep{mesasdk_linux,mesasdk_macos},
\code{matplotlib} \citep{hunter_2007_aa},
\code{NumPy} \citep{der_walt_2011_aa}, and \code{mesa\_reader} \citep{mesa_reader}.}


\bibliographystyle{aasjournal}

\end{document}